\documentclass[prl,reprint,superscriptaddress,amsmath,amssymb,aps,nofootinbib]{revtex4-1}

\usepackage{graphicx}
\usepackage{dcolumn}

\usepackage[english]{babel}
\usepackage[utf8]{inputenc}

\usepackage{enumitem} 
\usepackage{revsymb}
\usepackage{xcolor}
\usepackage{bbold}
\usepackage{bm}
\usepackage[colorlinks]{hyperref}

\hypersetup{
	colorlinks   = true,
	citecolor    = gray,
	linkcolor	 = blue,
	urlcolor = black
}

\usepackage{newfloat}

\usepackage{etoolbox}
\AtEndEnvironment{algorithm}{\noindent\hrulefill\par\nobreak\vskip-8pt}

\DeclareFloatingEnvironment[
fileext=loa,
listname=List of Algorithms,
name=ALGORITHM,
placement=tbhp,
]{algorithm}

\usepackage{amssymb}

\usepackage{array}
\newcolumntype{L}[1]{>{\raggedright\let\newline\\\arraybackslash\hspace{0pt}}m{#1}}
\newcolumntype{C}[1]{>{\centering\let\newline\\\arraybackslash\hspace{0pt}}m{#1}}
\newcolumntype{R}[1]{>{\raggedleft\let\newline\\\arraybackslash\hspace{0pt}}m{#1}}

\newcommand{\main}{\noindent\hspace*{10pt}\ignorespaces}
\newcommand{\bra}{\ensuremath{\langle\langle}}
\newcommand{\ket}{\ensuremath{\rangle\rangle}}

\makeatletter
\renewcommand{\fnum@figure}{FIG. \thefigure}
\makeatother

\begin{document}
	
	\title{Variational Quantum Monte Carlo Method with a Neural-Network
		Ansatz \\ for Open Quantum Systems}
	
	\author{Alexandra Nagy}
	\affiliation{Institute of Physics, Ecole Polytechnique Fédérale de Lausanne (EPFL), CH-1015, Lausanne, Switzerland}
	\author{Vincenzo Savona}
	\affiliation{Institute of Physics, Ecole Polytechnique Fédérale de Lausanne (EPFL), CH-1015, Lausanne, Switzerland}

	\begin{abstract}
		The possibility to simulate the properties of many-body open quantum systems with a large number of degrees of freedom is the premise to the solution of several outstanding problems in quantum science and quantum information. The challenge posed by this task lies in the complexity of the density matrix increasing exponentially with the system size. Here, we develop a variational method to efficiently simulate the non-equilibrium steady state of Markovian open quantum systems based on variational Monte Carlo and on a neural network representation of the density matrix. Thanks to the stochastic reconfiguration scheme, the application of the variational principle is translated into the actual integration of the quantum master equation. We test the effectiveness of the method by modeling the two-dimensional dissipative XYZ spin model on a lattice.
	\end{abstract}
	
	\maketitle
	
Open quantum systems have evolved into a major field of studies in recent years. Focus of these studies are the characterization of emergent phenomena and dissipative phase transitions \cite{Carusotto2013,Hartmann2016,Noh2017a,Bartolo2016,Biella2017,Biondi2017,Carmichael2015,Casteels2017,Casteels2016,Fink2017,Fink2018,Fitzpatrick2017,Foss-Feig2017,Kessler2012,Marino2016,Savona2017,Sieberer2013,Vicentini2018,Jin2016,Lee2013,Rota2018,Rota2017,Casteels2018}, as well as the ongoing debate about whether quantum computing schemes are still hard to simulate classically -- and thus achieve quantum supremacy -- when in presence of some degree of noise-induced decoherence \cite{Bremner2016,Gao2018,Harrow2017,Preskill2018}.

Assuming a Markovian interaction with the environment, the dynamics of open quantum systems is governed by the quantum master equation in Lindblad form \cite{Breuer2002}. Only few models within this description admit an analytical solution \cite{Prosen2011,Prosen2014}. The quest for efficient numerical methods to simulate the dynamics and the asymptotic steady state resulting from the Lindblad master equation, is a research field that is still in its infancy. Many recent tools have been developed following in the footsteps of well established numerical methods for the simulations of closed, Hamiltonian quantum systems. In particular, matrix-product state and tensor network schemes \cite{Cui2015,Kshetrimayum2017,Mascarenhas2015,Werner2016}, a real-space renormalization approach \cite{Finazzi2015}, cluster mean-field \cite{Jin2016}, and other ad-hoc approximation schemes \cite{Casteels2018,nagy_driven-dissipative_2018} have recently emerged.

A groundbreaking progress in the numerical simulation of both the ground state and the dynamics of closed quantum systems has recently been made with the introduction of the neural-network variational ansatz \cite{Cai2018,Carleo2018,Carleo2017,Freitas2018,Glasser2018,Nomura2017,Gao2017}, which efficiently represents highly correlated quantum states and whose parameters are easily optimized by means of the variational Monte Carlo (VMC) method. Recently, a self-adjoint and positive semi-definite parametrization of the density matrix, in terms of a neural network has been introduced \cite{Torlai2018}.

The steady state of an open quantum system can be characterized by a variational principle \cite{Cui2015,Jakob2003,Mascarenhas2015,Weimer2015}, whereby the dissipative part of the real-time dynamics, under quite general conditions, drives the system towards a unique steady state, in analogy to the imaginary-time Schrödinger equation that leads to the ground state of Hamiltonian systems.

In this Letter, we present a VMC approach to simulate the non-equilibrium steady state (NESS) of open quantum systems governed by the quantum master equation in Lindblad form. The density matrix is parametrized using a neural network ansatz \cite{Torlai2018} and parameters are varied using an extension of the stochastic reconfiguration method \cite{Sorella2007}, which is shown to approximate the real-time dynamics of the system. We apply the present VMC to study the steady-state properties of the dissipative XYZ spin model \cite{Jin2016,Lee2013,Rota2018,Rota2017,Casteels2018}, that displays a prototypical second-order dissipative phase transition. Thanks to the Monte-Carlo sampling of expectation values, this method holds promise for the efficient simulation of open quantum systems with a large number of degrees of freedom.\\
	
\textit{Dynamics of open quantum systems} -- The dynamics of the density matrix $\hat{\rho}$ of an open quantum system is governed by the quantum master equation which -- in case of Markovian coupling to the environment -- takes the Lindblad form

\begin{equation}\label{lindblad}
\frac{d\hat{\rho}}{d t} = -i[\hat{H}, \hat{\rho}] - \sum\limits_{i}^{}\frac{\gamma_i}{2}\left[\left\{\hat{F}_i^{\dagger}\hat{F}_i, 
\hat{\rho}\right\}-2\hat{F}_i\hat{\rho}\hat{F}_i^{\dagger}\right]\,,
\end{equation}

\noindent where the curly brackets denote the anti-commutator. The unitary part of the dynamics is generated by the term depending on the Hamiltonian $\hat{H}$, while $\hat{F}_i$ are the jump operators associated to the dissipative processes induced by the environment. The equation is typically expressed in terms of the Liouvillian superoperator as $d\hat{\rho}/d t = \mathcal{L}(\hat{\rho})$, whose formal solution is $\hat{\rho}(t)=e^{\mathcal{L}t}\hat{\rho}(0)$ ($t>0$). The existence and uniqueness of a NESS -- defined as $\hat\rho_{ss}=\lim_{t\to\infty}\hat\rho(t)$ -- satisfying
\begin{equation}\label{steady}
	 \mathcal{L}(\hat{\rho}_{ss})=0 \,,
\end{equation}
has been demonstrated under quite general assumptions \cite{Nigro2018,Minganti2018}, in particular for finite-size spin and boson lattices \cite{Nigro2018}.

The steady state can be computed as the long-time limit of the solution of the quantum master equation, or by directly solving the homogeoeous linear system \eqref{steady} with an additional condition on the trace of the density matrix. In both cases, the size of the problem is quantified by the square of the Hilbert space dimension, thus becoming computationally prohibitive already for a modest number of degrees of freedom. A promising route to the numerical computation of the NESS is provided by the variational principle. In cases where a unique steady state exists \cite{Nigro2018}, the NESS corresponds to the eigen-matrix of the Liouvillian super-operator $\mathcal{L}$ with zero eigenvalue \cite{Minganti2018}. As all other eigenvalues have strictly negative real part, the NESS can be formally derived as the matrix that maximizes the real part of the expectation value (computed in matrix space) of the Liouvillian.	\\

\textit{Neural Network Density Matrix} -- We assume that the Hilbert space of the system is spanned by the computational basis $|\boldsymbol{\sigma}\rangle$, where $\boldsymbol{\sigma}=(\sigma_1, \sigma_2, \dots, \sigma_N)$ labels the states of $N$ degrees of freedom that compose the system. Here and in what follows we will assume binary local degrees of freedom, with $\sigma_i=\{-1, 1\}$, which applies to the broad class of interacting spin-$1/2$ or qubit models. The density matrix in this basis is formally expressed as $\rho(\boldsymbol{\sigma},\boldsymbol{\eta})=\langle\boldsymbol{\sigma}|\hat{\rho}|\boldsymbol{\eta}\rangle$ in terms of the density operator $\hat{\rho}$. We denote a specific variational ansatz for the density matrix as $\rho_\chi(\boldsymbol{\sigma},\boldsymbol{\eta})$, where $\chi=(\chi_1,\,\chi_2,\ldots,\,\chi_{N_p})$ is a set of variational parameters. 

A neural network ansatz for a self-adjoint, positive semi-definite density matrix was recently introduced \cite{Torlai2018} in the specific form of a Restricted Boltzmann Machine (RBM). In a variational approach, RBMs present the significant advantage that the sum over the hidden-spin configurations can be carried out analytically, and the logarithmic derivatives with respect to the variational parameters admit simple expressions \cite{Carleo2017}. Here we briefly describe how this ansatz can be derived from simple considerations on the density matrix. A self-adjoint, positive semi-definite expression for the density matrix is

\begin{equation}\label{mixed}
	\rho_\chi(\boldsymbol{\sigma},\boldsymbol{\eta}) = \sum_{j=1}^{J} p_j(\chi)\cdot    \psi_j(\boldsymbol{\sigma},\chi)\psi_j^*(\boldsymbol{\eta},\chi)
\end{equation}

\noindent The states $\psi_j(\boldsymbol{\sigma},\chi)$ are not necessarily mutually orthogonal and the sum extends over $J$ states, with $J\leq d$ and $d=2^N$ is the dimension of the Hilbert space under study.

\begin{figure}[t!]
	\centering\includegraphics[width = 0.5\textwidth]{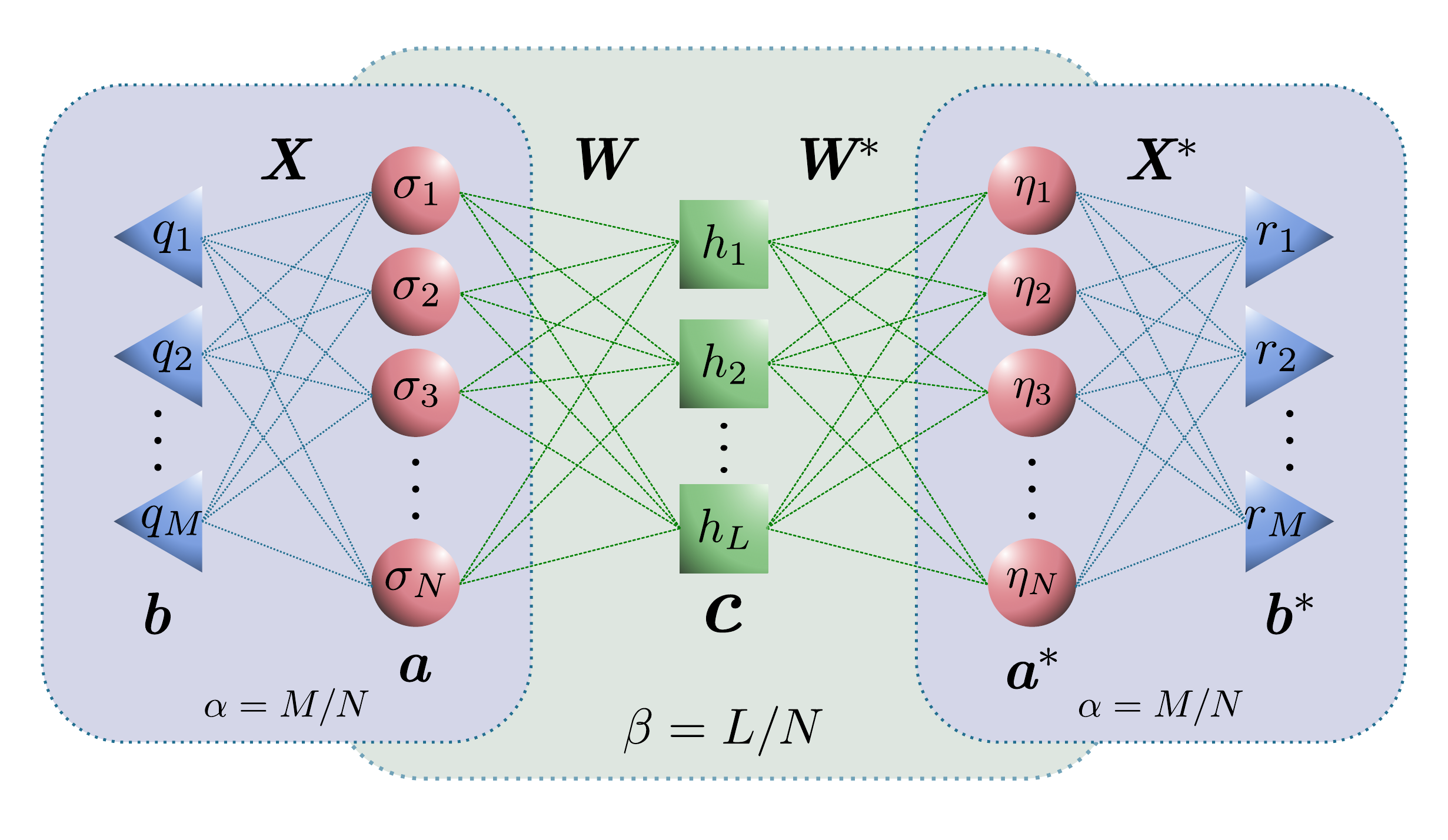}
	\caption{\label{fig:ansatz} Graphical representation of the neural network ansatz for the density matrix. The input states $|\boldsymbol{\sigma}\rangle, |\boldsymbol{\eta}\rangle$ are encoded in the visible layer, represented by circles. The hidden spins in the triangles encode the correlation between the physical spins in each state of the statistical mixture, while the hidden spins in the squares encode the mixture between the states. This structure is easily seen to coincide with a RBM, where the hidden layer is composed by the triangle and square nodes. }
\end{figure}

We start by introducing a RBM ansatz for each state $\psi_j(\boldsymbol{\sigma},\chi)$ entering expression \eqref{mixed}. A RBM is composed of two layers of binary valued nodes (see Fig.~\ref{fig:ansatz}): a visible layer for encoding the physical state and a hidden layer. Each node is associated with a bias ($a$- and $b$-parameters) and nodes in the different layers are connected via a set of weighted edges ($X$-parameters). For a large number of hidden nodes, this structure is known to describe quantum correlations efficiently \cite{Carleo2018,Glasser2018}.

In order to express the mixed structure in eq.~\eqref{mixed} as a single RBM, we embed an intermediate set of $L$ hidden nodes that are used to express the probabilities $p_j(\chi)$ in RBM form as $p_j(\chi)=\exp(\sum_{l}^{}c_lh_l)$, with $h_l=\pm1$ and $c_l\in \mathbb{R}$. To index the different states in the mixture accordingly, this new set of hidden nodes must also enter the RBM expression of the wave functions. When carrying out the sums over configurations of hidden nodes, the final expression for the RBM density matrix is \cite{supplemental}

\begin{equation}
\begin{split}
\rho_\chi(\boldsymbol{\sigma},\boldsymbol{\eta}) &=  8\exp\left(\sum_{i}^{}a_i\sigma_i\right)\exp\left(\sum_{i}^{}a_i^*\eta_i\right)\\
&\times\prod_{l=1}^{L}\cosh\left(c_l+\sum_{i}^{}W_{li}\sigma_i + \sum_{i}^{}W_{li}^*\eta_i\right) \\
&\times\prod_{m=1}^{M}\cosh\left(b_m+\sum_{i}^{}X_{mi}\sigma_i\right) \\ &\times\prod_{n=1}^{M}\cosh\left(b_n^*+\sum_{i}^{}X_{ni}^*\eta_i\right) \,.
\end{split}
\end{equation}

The RBM is sketched in Fig.~\ref{fig:ansatz}, and $\chi = \{a_i,b_m,X_{mi},c_l,W_{li}\}$ is the final set of parameters, which are assumed as complex valued with the exception of $c_l$ that must take real values. The representational power of the RBM is determined by the number of hidden nodes \cite{Leroux2008}. Here we set the densities of hidden nodes through the parameters $\alpha=M/N$, $\beta=L/N$, which measure the representational power of the RBM ansatz independently of the size of the spin lattice. When separately accounting for the real and imaginary parts of complex-valued parameters, the total number of computational parameters in the RBM ansatz is $N_p=N[(\alpha+\beta)(2N+1)+\alpha+2]$. In what follows, we will always assume $\alpha=\beta$ for simplicity.\\

\begin{figure}[t!]
	\centering\includegraphics[width = 0.5\textwidth]{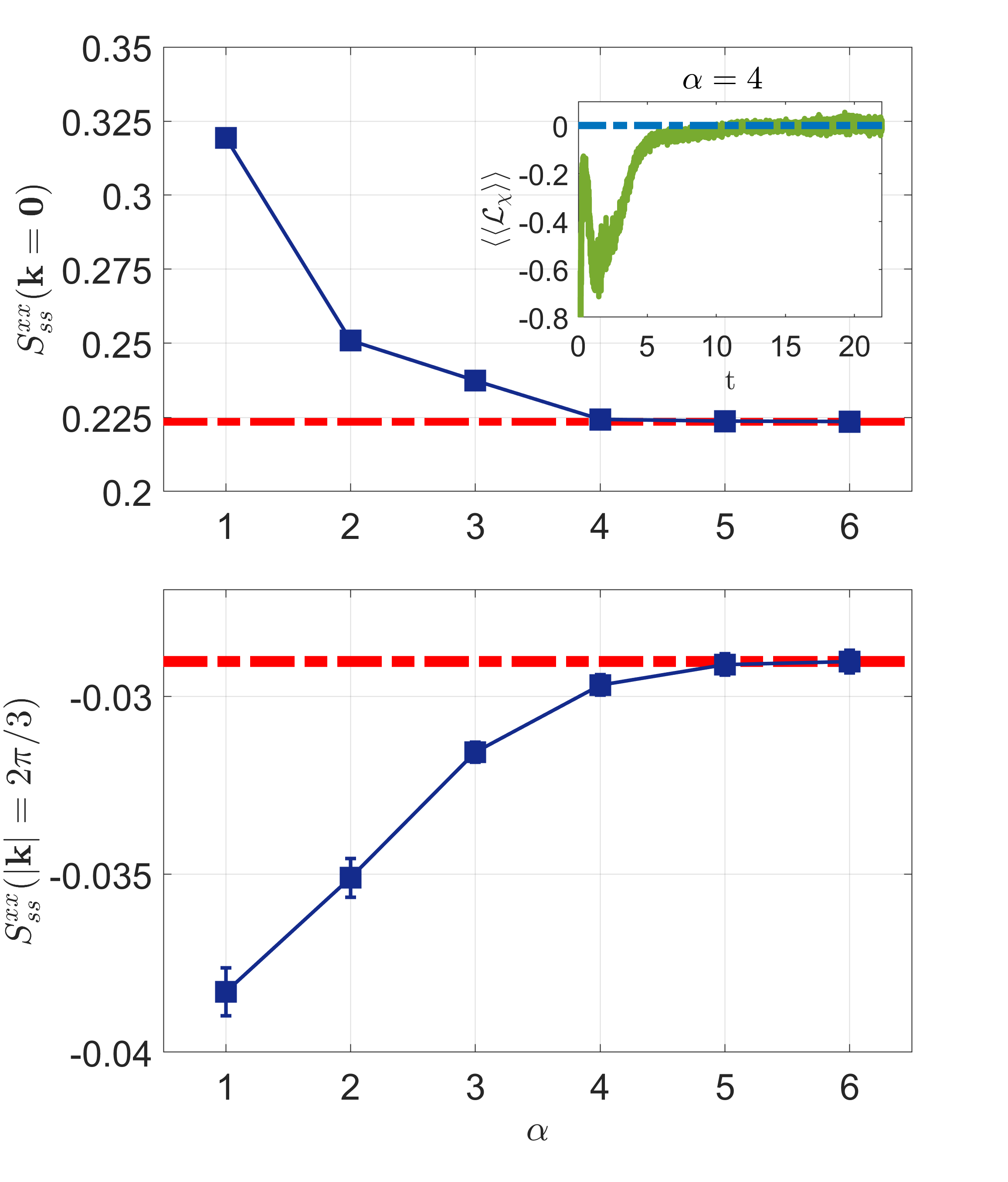}
	\caption{\label{fig:convergence} The steady-state spin structure factor $S_{ss}^{xx}(\mathrm{\textbf{k}})$ computed as a function of $\alpha=\beta$ for a 3$\times$3 lattice and $\mathrm{\textbf{k}=\textbf{0}}$ (upper panel) and $\mathrm{\textbf{k}=(2\pi/3,0)}$ (lower panel). The red dot-dashed line represents in both panels the exact result. The inset shows the evolution of $\bra\mathcal{L}_\chi\ket$ over the VMC run. Parameters: $J_x/\gamma=0.9$, $J_y/\gamma=1.2$, $J_z/\gamma = 1.0$.}
\end{figure}

\textit{Optimization} -- It is convenient to rewrite eq.~\eqref{steady} in a vectorized form by reshaping $\hat{\rho}$ into a column vector $|\rho\rangle$. Following \cite{Jakob2003}, $\mathcal{L}$ takes matrix form and the steady state density matrix fulfills $\langle \rho|\mathcal{L}|\rho\rangle=0$. Therefore the expectation value over the variational density matrix $\bra\mathcal{L}_\chi\ket=\langle \rho_\chi|\mathcal{L}|\rho_\chi\rangle/\langle \rho_\chi|\rho_\chi\rangle$ is a function of the variational parameters $\chi$. The parameter values that best approximate $\bra\mathcal{L}_\chi\ket=0$ can be found by means of various optimization procedures \cite{Torlai2018,Torlai2018a,Carleo2017,Nomura2017}. In this Letter, we choose to adopt the Stochastic Reconfiguration (SR) scheme by Sorella et al. \cite{Sorella2007} which we extend to open quantum systems. The parameters are initialized to a small random value and, at each iteration, they are updated as

\begin{equation}
	\chi(n+1)=\chi(n) + \nu\cdot S^{-1}(n)F(n)\,,
\end{equation}

\noindent where the learning rate $\nu$ is small enough to guarantee convergence. It can be shown \cite{supplemental} that the SR scheme induces, at each iteration, a variation in the parameters that best approximates the time evolution of the density matrix over a time step $\nu$. Here, we define the covariance matrix $S$, the vector of forces $F$, and the logarithmic derivatives $O$ as

\begin{equation}\label{sr}
	\begin{split}
		O_k(\boldsymbol{\sigma},\boldsymbol{\eta})&=\frac{1}{\rho_\chi(\boldsymbol{\sigma},\boldsymbol{\eta})}\cdot\frac{\partial \rho_\chi(\boldsymbol{\sigma},\boldsymbol{\eta})}{\partial \chi_k} \\
		F_k(n)&=\bra O_k^*\mathcal{L}\ket - \bra\mathcal{L}\ket\bra O_k^*\ket \\
		S_{kk'}(n)&=\bra O_k^*O_{k'}\ket - \bra O_{k}^*\ket\bra O_{k'}\ket \,,
	\end{split}
\end{equation}

\noindent where $k,k'=1,\,2,\ldots,\,N_p$. The notation $\bra \cdot \ket$ denotes the normalized expectation value taken over the variational density matrix $|\rho_\chi\rangle$, and the derivatives $O_k(\boldsymbol{\sigma},\boldsymbol{\eta})$ are taken as diagonal operators in these expectation values. We point out that, while the expression for $S$ in \eqref{sr} results in the VMC iterations following the real time evolution, minimization can be achieved by using any positive-definite covariance matrix. In particular, setting $S$ as the identity results in the steepest descent procedure. Since $S$ can be non-invertible, we apply an explicit regularization scheme, as introduced in \cite{Carleo2017}: 
$S_{kk'}^{\mathrm{reg}}=S_{kk'}+\lambda(n)\delta_{k,k'}S_{kk'}$, 
where $\lambda(n)=\max(\lambda_0b^n,\lambda_{\min})$. For the present calculations, they were set to $\lambda_0=100$, $b=0.998$ and $\lambda_{\min}=10^{-2}$.


\textit{Sampling} -- The various expectation values in \eqref{sr} must be evaluated at each iteration step. We evaluate these quantities stochastically over a Markov-chain of $N_{MH}$ configurations $(\boldsymbol{\sigma},\boldsymbol{\eta})$ sampling the square modulus of the density matrix $|\rho_\chi(\boldsymbol{\sigma},\boldsymbol{\eta})|^2$. For this we adopt the Metropolis-Hastings algorithm \cite{metropolis_equation_1953}. In the limit of $N_{MH}\rightarrow\infty$, the statistical error decays as $1/\sqrt{N_{MH}}$. Choosing an appropriate set of rules for the random walk is key to an efficient Monte Carlo sampling. Here we randomly choose each move among those allowed by the Liouvillian superoperator \cite{supplemental}.\\

\textit{Observables} -- Once the optimal parameter values have been determined, the expectation value of any quantum mechanical observable $\hat{\mathcal{O}}$ over the steady state can be expressed as

\begin{equation}
	\langle\hat{\mathcal{O}}\rangle=\mathrm{Tr}(\hat{\mathcal{O}}\hat{\rho}_\chi)=\sum_{\boldsymbol{\sigma},\boldsymbol{\eta}}^{}|\rho_\chi(\boldsymbol{\sigma},\boldsymbol{\eta})|^2\cdot\frac{\mathcal{O}(\boldsymbol{\eta},\boldsymbol{\sigma})}{\rho_\chi(\boldsymbol{\sigma},\boldsymbol{\eta})^*}\,,
\end{equation}

\noindent which can also be evaluated using the Metropolis-Hastings algorithm. For all the quantities considered here, the expectation values were additionally averaged over 100 sets of parameter values $\chi(n)$ chosen in the asymptotic region of the SR interation, in order to improve the statistical accuracy. The overall error in the sampled observables has, in addition to the contribution from the Metropolis-Hastings algorithms, a contribution from the SR scheme and a systematic contribution related to the representational power of the RBM ansatz, as measured by the $\alpha$ and $\beta$ parameters.\\

\begin{figure}[t!]
	\centering\includegraphics[width = 0.515\textwidth]{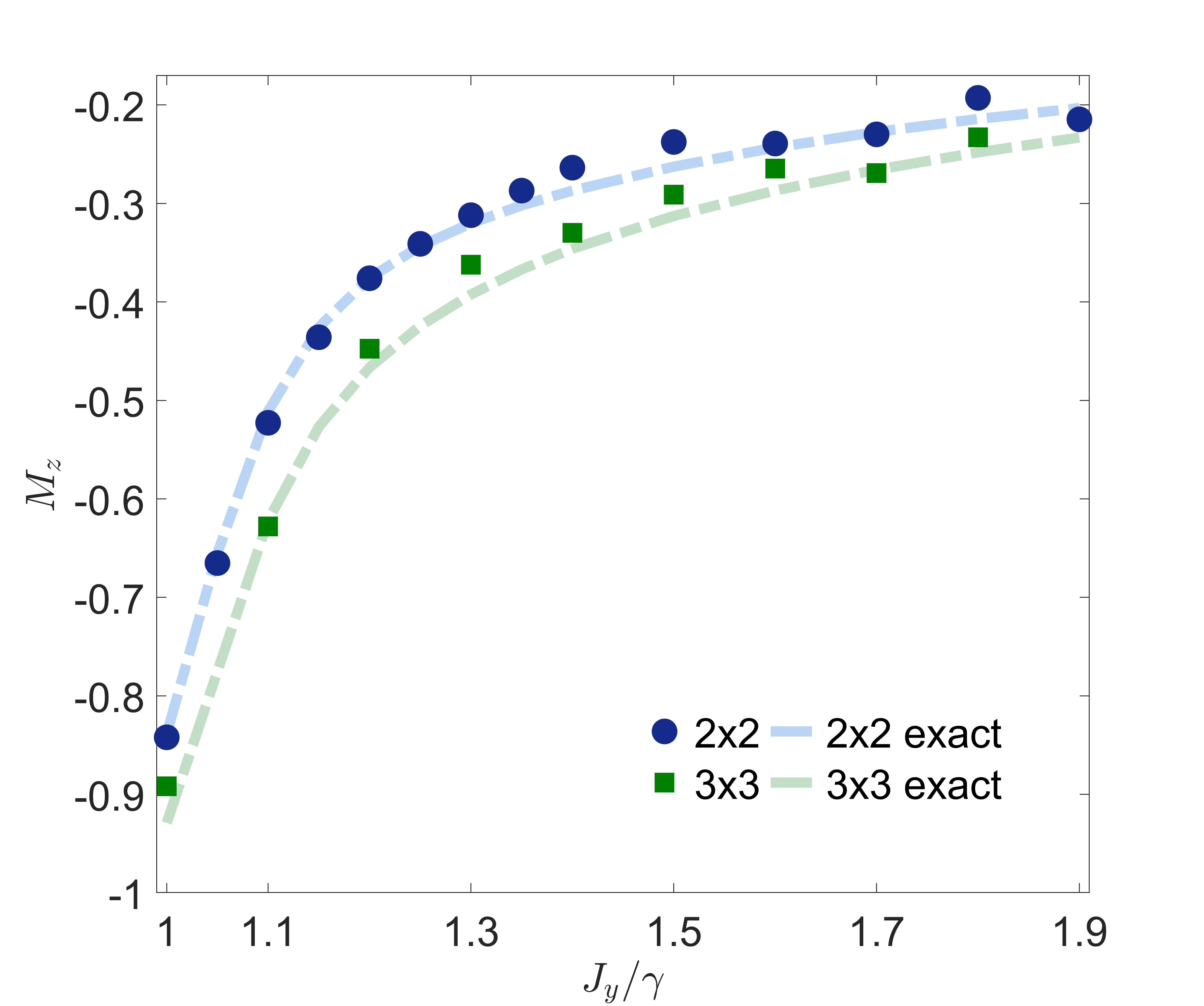}
	\caption{\label{fig:magnet} The magnetization $M_z$ computed as a function of the coupling $J_y/\gamma$. VMC and exact values are compared. Error bars, when not shown, are smaller than the symbol. Other parameters: $J_x/\gamma=0.9$, $J_z/\gamma = 1.0$, $\alpha=\beta=3$.}
\end{figure}

\textit{Computational cost} -- The number of floating point operations to evaluate Eq. (\ref{sr}) scales as $N_p^3$, if we assume that the number of Metropolis-Hastings steps $N_{MH}$ is set to roughly the number of parameters $N_p$, as in Ref. \cite{Carleo2017}. The Metropolis-Hastings procedure also scales with the number of connected states $N_c$, i.e. with the average number of nonzero elements in a column of the Liouvillian matrix. Finally, the efficiency of the whole procedure thus scales as $O(N_p^3+N_pN_c)$.

\textit{Results} -- To assess the effectiveness of the method, we study a spin-$1/2$ XYZ model on a two-dimensional lattice with periodic boundary condition. Each spin is subject to a dissipation process into the $|\sigma^z=-1\rangle$ state. This model has been already widely investigated and is known to display a dissipative phase transition between a paramegnetic and a ferromagnetic phase \cite{Jin2016,Lee2013,Rota2018,Rota2017,Casteels2018}. The Hamiltonian and the quantum master equation read ($\hbar=1$)

\begin{eqnarray}
\hat{H} = \sum\limits_{\langle i,j\rangle}^{}\left(J_x\hat{\sigma}_i^x\hat{\sigma}_j^x + J_y\hat{\sigma}_i^y\hat{\sigma}_j^y + J_z\hat{\sigma}_i^z\hat{\sigma}_j^z\right) \\
\frac{\mathrm{d}\hat{\rho}}{\mathrm{d}t} = -i[\hat{H}, \hat{\rho}] - \frac{\gamma}{2}\sum\limits_k^{}\left[\left\{\hat{\sigma}_k^+\hat{\sigma}_k^-, 
\hat{\rho}\right\}-2\hat{\sigma}_k^-\hat{\rho}\hat{\sigma}_k^+\right]
\end{eqnarray}

\noindent where $\hat{\sigma}_j^x$, $\hat{\sigma}_j^y$, $\hat{\sigma}_j^z$ are the Pauli matrices, $\hat\sigma^{\pm}_j=(\hat\sigma^x_j\pm i\hat\sigma^y_j)/2$, $J_\alpha$ are the coupling constants between nearest neighbour spins and $\gamma$ is the dissipation rate. The excitations in the system -- induced by the anisotropic spin coupling -- compete with the isotropic dissipative process, and this competition is at the origin of the dissipative phase transition  \cite{Jin2016,Lee2013,Rota2018,Rota2017,Casteels2018}. The effectiveness of the neural network ansatz is demonstrated by studying the system observables across a phase boundary. \\

In addition to the expectation value $\bra\mathcal{L}_\chi\ket$, we study the local magnetization 
\begin{equation}
M_z = \frac{1}{N}\sum_{i=1}^{N}\mathrm{Tr}(\hat{\rho}\hat{\sigma}_i^z) \,,
\end{equation}
and the steady-state structure factor
\begin{equation}
	S_{ss}^{xx}(\mathrm{\textbf{k}})=\frac{1}{N(N-1)}\sum_{\mathrm{\textbf{j}}\neq\mathrm{\textbf{l}}}^{}e^{-i\mathrm{\textbf{k}}(\mathrm{\textbf{j}}-\mathrm{\textbf{l}})}\langle\hat{\sigma}_{\mathrm{\textbf{j}}}^x\hat{\sigma}_{\mathrm{\textbf{l}}}^x\rangle \,,
\end{equation}
computed for the asymptotic steady state.

\begin{figure}[t!]
	\centering\includegraphics[width = 0.5\textwidth]{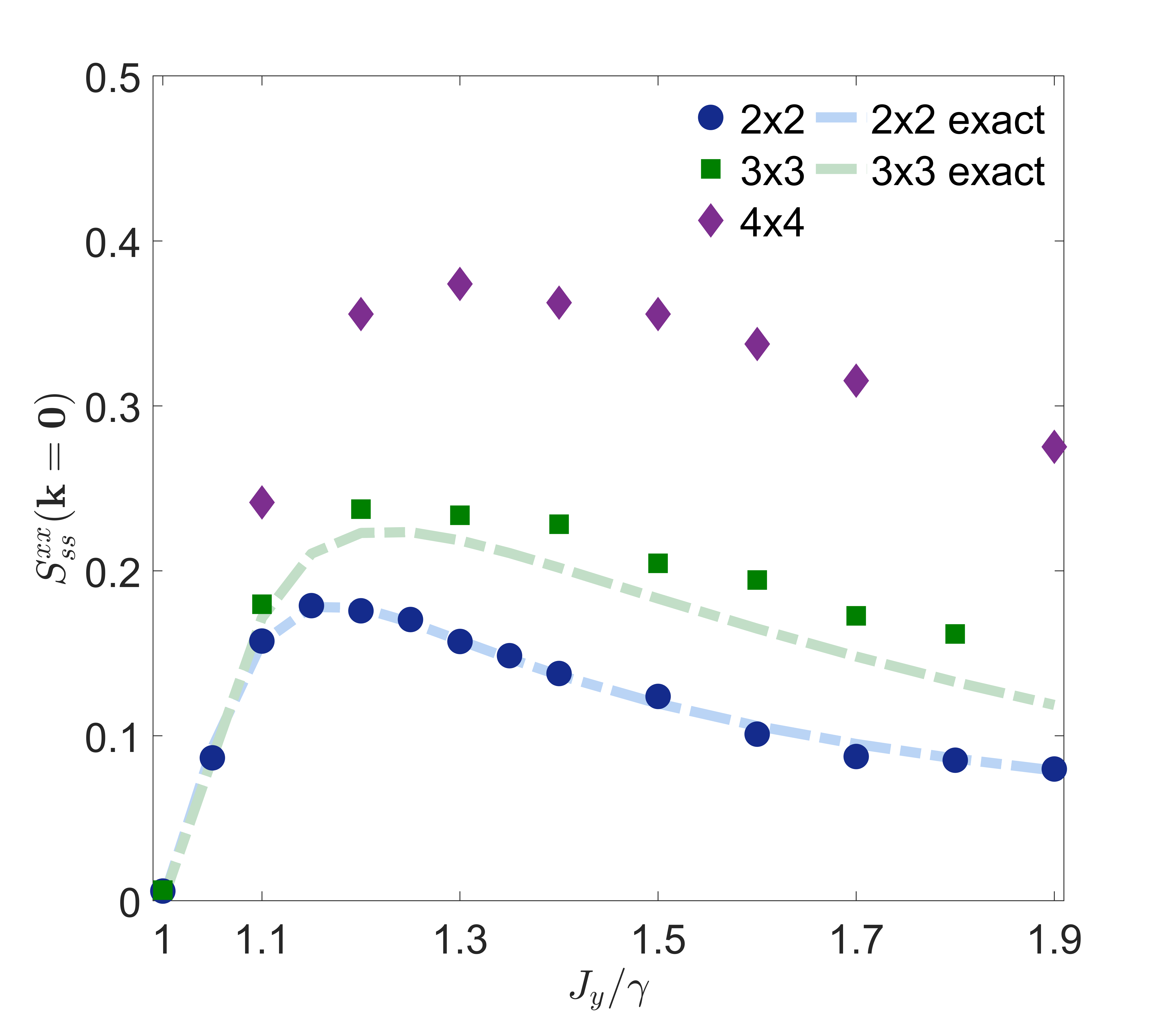}
	\caption{\label{fig:sxx} The steady-state spin structure factor $S_{ss}^{xx}(\mathrm{\textbf{k}}=\textbf{0})$ computed as a function of the coupling $J_y/\gamma$. VMC and exact values are compared. Other parameters: $J_x/\gamma=0.9$, $J_z/\gamma = 1.0$, $\alpha=\beta=3$.}
\end{figure}
\main Fig.~\ref{fig:convergence} shows the convergence of $S_{ss}^{xx}(\mathrm{\textbf{k}}=\textbf{0})$ and $S_{ss}^{xx}(\mathrm{\textbf{k}}=(2\pi/3,0))$ to the exact result for a $3\times 3$ lattice, as $\alpha=\beta$ are increased. The parameters $\chi$ are initialized randomly and updated at each VMC step according to the SR scheme \cite{supplemental}. The parameters of the model are chosen to lie in the vicinity of the dissipative phase transition, i.e. $J_x/\gamma=0.9$, $J_y/\gamma=1.2$, $J_z/\gamma = 1.0$. A clear convergence towards the exact value upon increasing $\alpha=\beta$ is found. The inset in Fig.~\ref{fig:convergence} shows the SR evolution of $\mathrm{Re}(\bra\mathcal{L}_\chi\ket)$ over a typical VMC run. The oscillations at early times are a feature of the unitary part of the dynamics in the quantum master equation. \\

\main In Fig.~\ref{fig:magnet} we display the magnetization as computed for different lattice sizes and as a function of the coupling parameter $J_y/\gamma$. For this choice of parameters, a para-to-ferromagnetic phase transition is expected to occur when increasing the coupling through the value $J_y \gtrapprox 1.04$ \cite{Rota2017,Rota2018}, while a second phase boundary between a ferromagnetic and a paramagnetic region has been predicted by cluster mean-field calculations at around $J_y\gtrapprox1.4$. For $2\times 2$ and $3\times 3$ lattices the VMC result agrees well with the exact calculation for a large enough number of variational parameters. 

\main In Fig.~\ref{fig:sxx} we display the spin structure factor $S_{ss}^{xx}(\mathrm{\textbf{k}}=\textbf{0})$ for the same parameters as in Fig.~\ref{fig:magnet}. The quantity $S_{ss}^{xx}(\mathrm{\textbf{k}}=\textbf{0})$ vanishes when in a paramagnetic phase, while it takes a finite value in the ferromagnetic region of the phase diagram. This behaviour is displayed both by the exact calculation for small lattices, and by the VMC data, in the vicinity of the phase boundary at $J_y \gtrapprox 1.04$. For values $J_y>1.4$ the system should become again paramagnetic in the thermodynamic limit of large lattices, but this feature was not displayed by the present data up to the largest lattice under study, in agreement also with recent stochastic Gutzwiller calculations \cite{Casteels2018}. \\
\main In all the calculations, special care was devoted to the choice of the SR time step $\nu$. The unitary part of the real-time dynamics generated by Eq.~\eqref{lindblad} makes the differential equation stiff, thus requiring to scale down $\nu$ appropriately as the system size -- and thus the spectral width of the differential operator -- is increased. A possible workaround would be to study an effective, purely dissipative dynamics using the super-operator $\mathcal{L}^\dagger\mathcal{L}$ as a generator. In the case of a unique steady state, this super-operator is self-adjoint and positive semi-defined, with the only null eigenvalue being associated to the steady-state solution. We argue that this effective dynamics would be more robust to the choice of the time step. The super-operator $\mathcal{L}^\dagger\mathcal{L}$ is however less sparse than $\mathcal{L}$ on the computational basis, calling for an efficient sampling scheme. \\
\main Existing numerical approaches to the simulation of the steady state of a Markovian open quantum system either require the full representation of the Hilbert space into memory, or rely on a properly chosen truncation of the Hilbert space to a relevant subspace. The present VMC approach is free of these two limitations, thanks to the stochastic evaluation of expectation values by means of the Metropolis-Hastings algorithm. The neural network ansatz in terms of a RBM is highly representative of quantum correlated statistical mixtures, while being simple to handle numerically. In cases with very strong quantum correlations, this ansatz could be extended to deep network representations, as was recently done in the case of Hamiltonian problems \cite{Carleo2018,Cai2018,Gao2017,Choo2018}. For some of these networks \cite{Choo2018}, the hidden degrees of freedom can still be summed analytically, as for RBMs. Neural network representations are not restricted to spin degrees of freedom and have been successfully adopted to represent bosonic many-body states efficiently \cite{Saito2017}. For these reasons, the present VMC approach may emerge as the election tool to numerically model open quantum systems, with considerable impact on the study of fundamental physics and on the modeling of near-term, noisy quantum information platforms \cite{Preskill2018}. 
\acknowledgments We are indebted to Giuseppe Carleo and Markus Holzmann for enlightening discussions. This work was supported by the Swiss National Science Foundation through Project No. 200021\_162357 and 200020\_185015.

\main While developing the present result we became aware of three related independent works that have been carried out in parallel \cite{hartmann_neural-network_2019,vicentini_variational_2019,yoshioka_constructing_2019}.

	\onecolumngrid
\newpage
\section{SUPPLEMENTAL MATERIAL}
\section*{Neural Network Representation of the Density Matrix}
\noindent In the main text we approximate the density matrix as a mapping $\rho_\chi$ with parameters $\chi$ which, given two input  states $|\boldsymbol{\sigma}\rangle$ and $|\boldsymbol{\eta}\rangle$, returns the matrix element $\langle\boldsymbol{\sigma}|\hat{\rho}_\chi|\boldsymbol{\eta}\rangle=\rho_\chi(\boldsymbol{\sigma},\boldsymbol{\eta})$. We describe quantum systems with $N$ degrees of freedom on a computational basis $\boldsymbol{\sigma}=(\sigma_1, \sigma_2, \dots, \sigma_N)$ assuming two-dimensional local Hilbert-spaces $\sigma_i=\{-1, 1\}$. This choice generally applies to spin-$1/2$ and quantum bits. A general expression, ensuring the positive semi-definite property of the density matrix, is 

\begin{equation}\label{sup_mixed}
	\rho_\chi(\boldsymbol{\sigma},\boldsymbol{\eta}) = \sum_{j=1}^{J} p_j(\chi)\cdot    \psi_j(\boldsymbol{\sigma},\chi)\psi_j^*(\boldsymbol{\eta},\chi)
\end{equation}

\noindent where $\chi$ are the parameters. The states $\psi_j(\boldsymbol{\sigma},\chi)$ are not necessarily mutually orthogonal and, together with the associated probabilities $1\ge p_j(\chi)\ge0$, are an expression of the statistical mixture described by $\hat{\rho}_\chi$. Here, the sum extends over $J$ states, with $J\leq d$ and $d=2^N$ is the dimension of the Hilbert-space under study. \\
Following Ref.~\cite{Carleo2017}, we may represent any wave function $\psi(\boldsymbol{\sigma})$ as a Restricted Boltzmann Machine (RBM),
\begin{equation}
	\psi(\boldsymbol{\sigma})=\sum_{\{q\}}^{}\exp\left(\sum_{i}^{}a_i\sigma_i+\sum_{m}^{}b_mq_m+\sum_{m,i}^{}q_m\sigma_iX_{mi}\right)
\end{equation}
\noindent where $q_m=\pm 1$ are a set of $M=\alpha\times N$ hidden spin variables and the leftmost sum runs over all possible hidden spin configurations $\{q\}$. The coupling and bias parameters $\{a_i,b_m,X_{mi}\}$ are, in general, complex-valued. In order to express the mixed state \eqref{sup_mixed} in terms of a single RBM, we introduce a second set of $L\beta\times N$ hidden spins that we use to represent the probabilities $p_j(\chi)$ in RBM form as $p_j(\chi)=\exp(\sum_{l}^{}c_lh_l)$, with $h_l=\pm1$ and $c_l\in \mathbb{R}$. This new set of hidden nodes should also enter in the expression of the wave functions, in order to index the different wave functions that make the mixed state. Hence, the full RBM form of the density matrix that we propose is
\begin{equation}
	\begin{split}
		\rho_\chi(\boldsymbol{\sigma},\boldsymbol{\eta}) &=  \sum_{\{h\}}^{}\sum_{\{q\}}^{}\sum_{\{r\}}^{}\exp\left(\sum_{l}^{}c_lh_l\right)\\
		&\times\exp\left(\sum_{i}^{}a_i\sigma_i+\sum_{m}^{}b_mq_m+\sum_{m,i}^{}q_m\sigma_iX_{mi} + \sum_{l,i}^{}h_l\sigma_iW_{li}\right) \\
		&\times\exp\left(\sum_{i}^{}a_i^*\eta_i+\sum_{n}^{}b_n^*r_n+\sum_{n,i}^{}r_n\eta_iX_{ni}^* + \sum_{l,i}^{}h_l\eta_iW_{li}^*\right)\,.
	\end{split}
\end{equation}
Since no intra-layer connection is allowed, the hidden variables can be explicitly traced out and the neural network density matrix reads as

\begin{equation}\label{finalsup}
	\begin{split}
		\rho_\chi(\boldsymbol{\sigma},\boldsymbol{\eta}) &=  8 \exp\left(\sum_{i}^{}a_i\sigma_i\right)\exp\left( \sum_{i}^{}a_i^*\eta_i\right)\\
		&\times\prod_{l=1}^{L}\cosh\left(c_l+\sum_{i}^{}W_{li}\sigma_i + \sum_{i}^{}W_{li}^*\eta_i\right) \\
		&\times\prod_{m=1}^{M}\cosh\left(b_m+\sum_{i}^{}X_{mi}\sigma_i\right) \\ &\times\prod_{n=1}^{M}\cosh\left(b_n^*+\sum_{i}^{}X_{ni}^*\eta_i\right) \,.
	\end{split}
\end{equation}

\noindent The variational parameters in this RBM are $\chi = \{a_i,b_m,X_{mi},c_l,W_{li}\}$ and, with the exception of the $c_l$'s, are all complex valued. When separately accounting for the real and imaginary parts of each parameter, the total number of real-valued computational parameters in the RBM ansatz is $N_p=N[(\alpha+\beta)(2N+1)+\alpha+2]$. We introduce the so called effective angles \cite{Carleo2017}
\begin{equation}
	\begin{split}
		\tilde{\theta}_l(\boldsymbol{\sigma},\boldsymbol{\eta}) &=c_l+\sum_{i}^{}W_{li}\sigma_i + \sum_{i}^{}W_{li}^*\eta_i \\
		\theta_m(\boldsymbol{\sigma}) &=b_m+\sum_{i}^{}X_{mi}\sigma_i \,,
	\end{split}
\end{equation}
\noindent finally obtaining
\begin{equation}\label{simp}
	\rho_\chi(\boldsymbol{\sigma},\boldsymbol{\eta}) = 8\exp\left(\sum_{i}^{}a_i\sigma_i\right)\exp\left( \sum_{i}^{}a_i^*\eta_i\right)
	\prod_{l=1}^{L}\prod_{m=1}^{M}\prod_{n=1}^{M}\cosh\tilde{\theta}_l(\boldsymbol{\sigma},\boldsymbol{\eta})\cosh\theta_m(\boldsymbol{\sigma})\cosh\theta_n^*(\boldsymbol{\eta}) \,.
\end{equation}

\noindent In order to find the best set of parameters to describe the steady state, we choose to use the Stochastic Reconfiguration (SR) algorithm. This requires the calculation of the logarithmic derivatives of $\rho_\chi(\boldsymbol{\sigma},\boldsymbol{\eta})$ with respect to the real and imaginary parts of all variational parameters. Given expression~\eqref{simp}, these are written in a rather compact form as 
\begin{equation}\label{final}
	\begin{split}
		\frac{1}{\rho_\chi(\boldsymbol{\sigma},\boldsymbol{\eta})}\cdot\frac{\partial \rho_\chi(\boldsymbol{\sigma},\boldsymbol{\eta})}{\partial \Re(a_k)} &=  \sigma_k +\eta_k \\
		\frac{1}{\rho_\chi(\boldsymbol{\sigma},\boldsymbol{\eta})}\cdot\frac{\partial \rho_\chi(\boldsymbol{\sigma},\boldsymbol{\eta})}{\partial \Im(a_k)} &=  i(\sigma_k -\eta_k) \\
		\frac{1}{\rho_\chi(\boldsymbol{\sigma},\boldsymbol{\eta})}\cdot\frac{\partial \rho_\chi(\boldsymbol{\sigma},\boldsymbol{\eta})}{\partial \Re(b_k)} &= \tanh\left(\theta_k(\boldsymbol{\sigma})\right)+\tanh\left(\theta_k^*(\boldsymbol{\eta})\right)  \\
		\frac{1}{\rho_\chi(\boldsymbol{\sigma},\boldsymbol{\eta})}\cdot\frac{\partial \rho_\chi(\boldsymbol{\sigma},\boldsymbol{\eta})}{\partial \Im(b_k)} &=i\left[ \tanh\left(\theta_k(\boldsymbol{\sigma})\right)-\tanh\left(\theta_k^*(\boldsymbol{\eta})\right)\right]  \\
		\frac{1}{\rho_\chi(\boldsymbol{\sigma},\boldsymbol{\eta})}\cdot\frac{\partial \rho_\chi(\boldsymbol{\sigma},\boldsymbol{\eta})}{\partial \Re(X_{kl})} &= \tanh\left(\theta_k(\boldsymbol{\sigma})\right)\sigma_l+\tanh\left(\theta_k^*(\boldsymbol{\eta})\right)\eta_l  \\
		\frac{1}{\rho_\chi(\boldsymbol{\sigma},\boldsymbol{\eta})}\cdot\frac{\partial \rho_\chi(\boldsymbol{\sigma},\boldsymbol{\eta})}{\partial \Im(X_{kl})} &=i\left[ \tanh\left(\theta_k(\boldsymbol{\sigma})\right)\sigma_l-\tanh\left(\theta_k^*(\boldsymbol{\eta})\right)\eta_l\right]  \\
		\frac{1}{\rho_\chi(\boldsymbol{\sigma},\boldsymbol{\eta})}\cdot\frac{\partial \rho_\chi(\boldsymbol{\sigma},\boldsymbol{\eta})}{\partial c_k} &=  \tanh\left(\tilde{\theta}_l(\boldsymbol{\sigma},\boldsymbol{\eta})\right) \\
		\frac{1}{\rho_\chi(\boldsymbol{\sigma},\boldsymbol{\eta})}\cdot\frac{\partial \rho_\chi(\boldsymbol{\sigma},\boldsymbol{\eta})}{\partial \Re(W_{kl})} &=  \tanh\left(\tilde{\theta}_l(\boldsymbol{\sigma},\boldsymbol{\eta})\right)(\sigma_l+\eta_l) \\
		\frac{1}{\rho_\chi(\boldsymbol{\sigma},\boldsymbol{\eta})}\cdot\frac{\partial \rho_\chi(\boldsymbol{\sigma},\boldsymbol{\eta})}{\partial \Im(W_{kl})} &= i \tanh\left(\tilde{\theta}_l(\boldsymbol{\sigma},\boldsymbol{\eta})\right)(\sigma_l-\eta_l) 
	\end{split}
\end{equation}

\vspace*{0.5cm}
\section*{Stochastic Reconfiguration for Open Quantum Systems}
\noindent We approximate the density matrix with a neural network ansatz $\rho_\chi$, in order to find the best representation of the steady state. In the following, we use the vectorized formalism of the Liouville-von-Neumann master equation by reshaping $\hat{\rho}$ into a column vector $|\rho\rangle$ and the Liouvillian superoperator into a matrix $\mathcal{L}$ of dimension $d^2\times d^2$ where $d$ is the dimension of the Hilbert-space under study.
In order to find the optimal set of parameters, we choose to extend the SR method \cite{Sorella2007} to open quantum systems.\\
In the SR optimization the variational parameters are changed at each iteration step by
\begin{equation}
	\chi_k'=\chi_k + \delta\chi_k\,.
\end{equation}
\noindent After a small perturbation in linear approximation the density matrix reads as 
\begin{equation}\label{lin}
	|\rho_{\chi'}\rangle\simeq\sum_{k}^{}\delta\chi_kO_k|\rho_\chi\rangle\,,
\end{equation}
\noindent where the diagonal operators $O_k$ are defined for any configuration $|\boldsymbol{\sigma},\boldsymbol{\eta}\rangle=|x\rangle$ as the variational derivative with respect to the $k$-th parameter
\begin{equation}
	O_k(x)=O_k(\boldsymbol{\sigma},\boldsymbol{\eta})=\frac{1}{\rho_\chi(\boldsymbol{\sigma},\boldsymbol{\eta})}\cdot\frac{\partial \rho_\chi(\boldsymbol{\sigma},\boldsymbol{\eta})}{\partial \chi_k} \,.
\end{equation}

\noindent Open quantum systems evolve under a one-parameters semigroup dictated by the Liouvillian superoperator as 
\begin{equation}\label{power}
	|\bar{\rho}_\chi\rangle=e^{\mathcal{L}t}|\rho_\chi\rangle\simeq(\mathbb{1}+\mathcal{L})|\rho_\chi\rangle\,.
\end{equation}

\noindent Under general assumptions, this dynamics asymptotically converges to the non-equilibrium steady state. We can enforce the real-time dynamics generated by the Liouvillian within the SR scheme. The method can be then interpreted as a real time evolution in the variational subspace. \\
To achieve this goal, we equate eq.~\eqref{lin} and~\eqref{power} in the subspace spanned by the vectors $\{O_k|\rho_\chi\rangle\}$, thus obtaining
\begin{equation}
	\delta\chi_k=\nu\sum_{k'}^{}S_{kk'}^{-1}F_{k'}\,,
\end{equation}
\noindent where $\nu$ is small enough to guarantee convergence and we introduced the generalized forces and the covariance matrix for a given set of variational parameters at iteration step $n$
\begin{equation}\label{sr_sup}
	\begin{split}
		F_k(n)&=\bra O_k^*\mathcal{L}\ket - \bra\mathcal{L}\ket\bra O_k^*\ket \\
		S_{kk'}(n)&=\bra O_k^*O_{k'}\ket - \bra O_{k}^*\ket\bra O_{k'}\ket \,.
	\end{split}
\end{equation}
\noindent Here, $\bra \cdot \ket$ denotes the normalized expectational value taken over the variational density matrix $|\rho_\chi\rangle$. The covariance matrix can be any positive-definite matrix. Choosing it as the identity will correspond to the steepest descent optimization method. Since $S$ can be non-invertible we apply an explicit regularization scheme, as introduced in \cite{Carleo2017}
\begin{equation}
	S_{kk'}^{\mathrm{reg}}=S_{kk'}+\lambda(n)\delta_{k,k'}S_{kk'}\,.
\end{equation}
\noindent Here, the $\lambda(n)$ parameter is a function of the iteration step $n$ and decays with $\lambda(n)=\max(\lambda_0b^n,\lambda_{\min})$. For the present calculations, they were set to $\lambda_0=100$, $b=0.998$ and $\lambda_{\min}=10^{-2}$.\\
\section*{Stochastic Sampling}
\noindent For the SR optimization scheme, one must evaluate~\eqref{sr_sup} at each iteration step. In order to perform a stochastic sampling of these quantities, we write the general forces and the covariance matrix into a suitable form as follows
\begin{equation}\label{metro}
	\begin{split}
		F_k(n)&= \sum_{x}^{}|\langle x|\rho_\chi\rangle|^2\cdot\left(\frac{\partial \ln\langle x|\rho_\chi\rangle}{\partial \chi_k}\right)^*\frac{\langle x|\mathcal{L}|\rho_\chi\rangle}{\langle x|\rho_\chi\rangle} \\
		&- \sum_{x}^{}|\langle x|\rho_\chi\rangle|^2\cdot\frac{\langle x|\mathcal{L}|\rho_\chi\rangle}{\langle x|\rho_\chi\rangle}\sum_{x'}^{}|\langle x'|\rho_\chi\rangle|^2\cdot\left(\frac{\partial \ln\langle x'|\rho_\chi\rangle}{\partial \chi_k}\right)^*\\
		S_{kk'}(n)&=\sum_{x}^{}|\langle x|\rho_\chi\rangle|^2\cdot\left(\frac{\partial \ln\langle x|\rho_\chi\rangle}{\partial \chi_k}\right)^*\left(\frac{\partial \ln\langle x|\rho_\chi\rangle}{\partial \chi_{k'}}\right) \\
		&- \sum_{x}^{}|\langle x|\rho_\chi\rangle|^2\cdot\left(\frac{\partial \ln\langle x|\rho_\chi\rangle}{\partial \chi_k}\right)^*\sum_{x'}^{}|\langle x'|\rho_\chi\rangle|^2\cdot\left(\frac{\partial \ln\langle x'|\rho_\chi\rangle}{\partial \chi_{k'}}\right) \,,
	\end{split}
\end{equation}

\noindent where $|x\rangle=|\boldsymbol{\sigma},\boldsymbol{\eta}\rangle$ are the basis elements spanning the space of the vectorized density matrix. Therefore, for a given $\chi$ we can generate a Markov-chain of $N_{MH}$ many-body configurations $|x_1\rangle \rightarrow |x_2\rangle \rightarrow \dots \rightarrow |x_{N_{MH}}\rangle$ sampling the square modulus of the density matrix elements $|\langle x|\rho_\chi\rangle|^2=|\rho_\chi(\boldsymbol{\sigma},\boldsymbol{\eta})|^2$ via a standard Metropolis-Hastings algorithm \cite{metropolis_equation_1953}. The configurations are generated based on a proposition scheme, and accepted according to the probability
\begin{equation}
	\mathcal{A}(|x_{y}\rangle\rightarrow |x_z\rangle) = \min\left(1,\frac{|\langle x_z|\rho_\chi\rangle|^2\cdot\mathcal{P}(y|z)}{|\langle x_y|\rho_\chi\rangle|^2\cdot\mathcal{P}(z|y)}\right) \,.
\end{equation}
\noindent where $\mathcal{P}(z|y)$ is the conditional probability of proposing a state $z$ given $y$. Our move generation is based on the transitions dictated by the Liouvillian superoperator which occur by applying $\mathcal{L}|x\rangle$. Doing so decreases the required number of thermalization and sampling steps by promoting, in the random walk, the configurations with the highest importance. However, in case of a close-to-pure steady-state, a handful of density matrix elements have magnitude dominating the others. This causes the random walk to perform poorly, as some regions in configuration space become scarcely accessible. Hence, we introduced additional moves which occur with a low probability and reach a configuration by randomly flipping one spin in both $\boldsymbol{\sigma}$ and $\boldsymbol{\eta}$. The propositions always need to obey detailed balance. \\
For sampling $|\rho_\chi(\boldsymbol{\sigma},\boldsymbol{\eta})|^2$ in the case of a two-dimensional spin lattice model, we introduce the following possible moves on the configuration space spanned by $|\boldsymbol{\sigma},\boldsymbol{\eta}\rangle$
\begin{enumerate}
	\item \textit{Column-hopping in $\boldsymbol{\sigma}$}: a site $\sigma_j$ and its right neighbour are flipped. This move accounts for single spin hopping as well as double excitation or loss.
	
	\item \textit{Row-hopping in $\boldsymbol{\sigma}$}: a site $\sigma_j$ and its down neighbour are flipped.
	
	\item \textit{Column-hopping in $\boldsymbol{\eta}$}: a site $\eta_j$ and its right neighbour are flipped.
	
	\item \textit{Row-hopping in $\boldsymbol{\eta}$}: a site $\eta_j$ and its down neighbour are flipped.
	
	\item \textit{Excitation in $\boldsymbol{\sigma}$}: a site $\sigma_j$ is flipped. This move corresponds to the effect of having an external field in the direction perpendicular to the quantization axis.
	
	\item \textit{Excitation in $\boldsymbol{\eta}$}: a site $\eta_j$ is flipped.
	
	\item \textit{Dissipator}: sites $\sigma_j=\eta_j$ are flipped with an asymmetrical transition ratio. The dissipative moves are always proposed, while excitation are only proposed with ten-percent probability. 
	
	\item\textit{Jumper:} two randomly chosen spins $\sigma_l,\eta_m$ are flipped.
	
\end{enumerate}

\noindent For the move proposition we choose one of the actions with uniform probability.

\section*{Detail on numerical approach}

The VMC code was written in Python. The Metropolis Hastings algorithm was efficiently parallelized by splitting the Markov chain into several independent chains that were run in parallel using MVAPICH2. Both a CPU and a GPU version of the code were developed. GPU calculations allowed to significantly speed up the update of the $O_k$ expectation values, as well as the iterative solution of the linear system in the SR, using the MINRES-QLP algorithm \cite{choi_minres-qlp:_2011,liu_contribute_2019} which can correctly handle the case of a singular matrix. The GPU version brought considerable advantage over CPU when the number of variational parameters was larger than 1000. In the Metropolis-Hastings sampling we set the number of {\em accepted} moves instead of the total number of moves. This is necessary as, for some choices of the physical parameters, the density matrix is highly concentrated on a few matrix elements and consequently MH moves are seldom accepted. The number of accepted moves was set to be roughly twice the number of variational parameters in all VMC runs.


\begin{thebibliography}{58}%
	\makeatletter
	\providecommand \@ifxundefined [1]{%
		\@ifx{#1\undefined}
	}%
	\providecommand \@ifnum [1]{%
		\ifnum #1\expandafter \@firstoftwo
		\else \expandafter \@secondoftwo
		\fi
	}%
	\providecommand \@ifx [1]{%
		\ifx #1\expandafter \@firstoftwo
		\else \expandafter \@secondoftwo
		\fi
	}%
	\providecommand \natexlab [1]{#1}%
	\providecommand \enquote  [1]{``#1''}%
	\providecommand \bibnamefont  [1]{#1}%
	\providecommand \bibfnamefont [1]{#1}%
	\providecommand \citenamefont [1]{#1}%
	\providecommand \href@noop [0]{\@secondoftwo}%
	\providecommand \href [0]{\begingroup \@sanitize@url \@href}%
	\providecommand \@href[1]{\@@startlink{#1}\@@href}%
	\providecommand \@@href[1]{\endgroup#1\@@endlink}%
	\providecommand \@sanitize@url [0]{\catcode `\\12\catcode `\$12\catcode
		`\&12\catcode `\#12\catcode `\^12\catcode `\_12\catcode `\%12\relax}%
	\providecommand \@@startlink[1]{}%
	\providecommand \@@endlink[0]{}%
	\providecommand \url  [0]{\begingroup\@sanitize@url \@url }%
	\providecommand \@url [1]{\endgroup\@href {#1}{\urlprefix }}%
	\providecommand \urlprefix  [0]{URL }%
	\providecommand \Eprint [0]{\href }%
	\providecommand \doibase [0]{http://dx.doi.org/}%
	\providecommand \selectlanguage [0]{\@gobble}%
	\providecommand \bibinfo  [0]{\@secondoftwo}%
	\providecommand \bibfield  [0]{\@secondoftwo}%
	\providecommand \translation [1]{[#1]}%
	\providecommand \BibitemOpen [0]{}%
	\providecommand \bibitemStop [0]{}%
	\providecommand \bibitemNoStop [0]{.\EOS\space}%
	\providecommand \EOS [0]{\spacefactor3000\relax}%
	\providecommand \BibitemShut  [1]{\csname bibitem#1\endcsname}%
	\let\auto@bib@innerbib\@empty
	\bibitem [{\citenamefont {Carusotto}\ and\ \citenamefont
		{Ciuti}(2013)}]{Carusotto2013}%
	\BibitemOpen
	\bibfield  {author} {\bibinfo {author} {\bibfnamefont {I.}~\bibnamefont
			{Carusotto}}\ and\ \bibinfo {author} {\bibfnamefont {C.}~\bibnamefont
			{Ciuti}},\ }\href {\doibase 10.1103/RevModPhys.85.299} {\bibfield  {journal}
		{\bibinfo  {journal} {Reviews of Modern Physics}\ }\textbf {\bibinfo {volume}
			{85}},\ \bibinfo {pages} {299} (\bibinfo {year} {2013})}\BibitemShut
	{NoStop}%
	\bibitem [{\citenamefont {Hartmann}(2016)}]{Hartmann2016}%
	\BibitemOpen
	\bibfield  {author} {\bibinfo {author} {\bibfnamefont {M.~J.}\ \bibnamefont
			{Hartmann}},\ }\href {http://stacks.iop.org/2040-8986/18/i=10/a=104005}
	{\bibfield  {journal} {\bibinfo  {journal} {Journal of Optics}\ }\textbf
		{\bibinfo {volume} {18}},\ \bibinfo {pages} {104005} (\bibinfo {year}
		{2016})}\BibitemShut {NoStop}%
	\bibitem [{\citenamefont {Noh}\ and\ \citenamefont
		{Angelakis}(2017)}]{Noh2017a}%
	\BibitemOpen
	\bibfield  {author} {\bibinfo {author} {\bibfnamefont {C.}~\bibnamefont
			{Noh}}\ and\ \bibinfo {author} {\bibfnamefont {D.~G.}\ \bibnamefont
			{Angelakis}},\ }\href {\doibase 10.1088/0034-4885/80/1/016401} {\bibfield
		{journal} {\bibinfo  {journal} {Reports on Progress in Physics}\ }\textbf
		{\bibinfo {volume} {80}},\ \bibinfo {pages} {016401} (\bibinfo {year}
		{2017})}\BibitemShut {NoStop}%
	\bibitem [{\citenamefont {Bartolo}\ \emph {et~al.}(2016)\citenamefont
		{Bartolo}, \citenamefont {Minganti}, \citenamefont {Casteels},\ and\
		\citenamefont {Ciuti}}]{Bartolo2016}%
	\BibitemOpen
	\bibfield  {author} {\bibinfo {author} {\bibfnamefont {N.}~\bibnamefont
			{Bartolo}}, \bibinfo {author} {\bibfnamefont {F.}~\bibnamefont {Minganti}},
		\bibinfo {author} {\bibfnamefont {W.}~\bibnamefont {Casteels}}, \ and\
		\bibinfo {author} {\bibfnamefont {C.}~\bibnamefont {Ciuti}},\ }\href
	{\doibase 10.1103/PhysRevA.94.033841} {\bibfield  {journal} {\bibinfo
			{journal} {Phys. Rev. A}\ }\textbf {\bibinfo {volume} {94}},\ \bibinfo
		{pages} {033841} (\bibinfo {year} {2016})}\BibitemShut {NoStop}%
	\bibitem [{\citenamefont {Biella}\ \emph {et~al.}(2017)\citenamefont {Biella},
		\citenamefont {Storme}, \citenamefont {Lebreuilly}, \citenamefont {Rossini},
		\citenamefont {Fazio}, \citenamefont {Carusotto},\ and\ \citenamefont
		{Ciuti}}]{Biella2017}%
	\BibitemOpen
	\bibfield  {author} {\bibinfo {author} {\bibfnamefont {A.}~\bibnamefont
			{Biella}}, \bibinfo {author} {\bibfnamefont {F.}~\bibnamefont {Storme}},
		\bibinfo {author} {\bibfnamefont {J.}~\bibnamefont {Lebreuilly}}, \bibinfo
		{author} {\bibfnamefont {D.}~\bibnamefont {Rossini}}, \bibinfo {author}
		{\bibfnamefont {R.}~\bibnamefont {Fazio}}, \bibinfo {author} {\bibfnamefont
			{I.}~\bibnamefont {Carusotto}}, \ and\ \bibinfo {author} {\bibfnamefont
			{C.}~\bibnamefont {Ciuti}},\ }\href {\doibase 10.1103/PhysRevA.96.023839}
	{\bibfield  {journal} {\bibinfo  {journal} {Phys. Rev. A}\ }\textbf {\bibinfo
			{volume} {96}},\ \bibinfo {pages} {023839} (\bibinfo {year}
		{2017})}\BibitemShut {NoStop}%
	\bibitem [{\citenamefont {Biondi}\ \emph {et~al.}(2017)\citenamefont {Biondi},
		\citenamefont {Blatter}, \citenamefont {T\"ureci},\ and\ \citenamefont
		{Schmidt}}]{Biondi2017}%
	\BibitemOpen
	\bibfield  {author} {\bibinfo {author} {\bibfnamefont {M.}~\bibnamefont
			{Biondi}}, \bibinfo {author} {\bibfnamefont {G.}~\bibnamefont {Blatter}},
		\bibinfo {author} {\bibfnamefont {H.~E.}\ \bibnamefont {T\"ureci}}, \ and\
		\bibinfo {author} {\bibfnamefont {S.}~\bibnamefont {Schmidt}},\ }\href
	{\doibase 10.1103/PhysRevA.96.043809} {\bibfield  {journal} {\bibinfo
			{journal} {Phys. Rev. A}\ }\textbf {\bibinfo {volume} {96}},\ \bibinfo
		{pages} {043809} (\bibinfo {year} {2017})}\BibitemShut {NoStop}%
	\bibitem [{\citenamefont {Carmichael}(2015)}]{Carmichael2015}%
	\BibitemOpen
	\bibfield  {author} {\bibinfo {author} {\bibfnamefont {H.~J.}\ \bibnamefont
			{Carmichael}},\ }\href {\doibase 10.1103/PhysRevX.5.031028} {\bibfield
		{journal} {\bibinfo  {journal} {Phys. Rev. X}\ }\textbf {\bibinfo {volume}
			{5}},\ \bibinfo {pages} {031028} (\bibinfo {year} {2015})}\BibitemShut
	{NoStop}%
	\bibitem [{\citenamefont {Casteels}\ \emph {et~al.}(2017)\citenamefont
		{Casteels}, \citenamefont {Fazio},\ and\ \citenamefont
		{Ciuti}}]{Casteels2017}%
	\BibitemOpen
	\bibfield  {author} {\bibinfo {author} {\bibfnamefont {W.}~\bibnamefont
			{Casteels}}, \bibinfo {author} {\bibfnamefont {R.}~\bibnamefont {Fazio}}, \
		and\ \bibinfo {author} {\bibfnamefont {C.}~\bibnamefont {Ciuti}},\ }\href
	{\doibase 10.1103/PhysRevA.95.012128} {\bibfield  {journal} {\bibinfo
			{journal} {Phys. Rev. A}\ }\textbf {\bibinfo {volume} {95}},\ \bibinfo
		{pages} {012128} (\bibinfo {year} {2017})}\BibitemShut {NoStop}%
	\bibitem [{\citenamefont {Casteels}\ \emph {et~al.}(2016)\citenamefont
		{Casteels}, \citenamefont {Storme}, \citenamefont {Le~Boit\'e},\ and\
		\citenamefont {Ciuti}}]{Casteels2016}%
	\BibitemOpen
	\bibfield  {author} {\bibinfo {author} {\bibfnamefont {W.}~\bibnamefont
			{Casteels}}, \bibinfo {author} {\bibfnamefont {F.}~\bibnamefont {Storme}},
		\bibinfo {author} {\bibfnamefont {A.}~\bibnamefont {Le~Boit\'e}}, \ and\
		\bibinfo {author} {\bibfnamefont {C.}~\bibnamefont {Ciuti}},\ }\href
	{\doibase 10.1103/PhysRevA.93.033824} {\bibfield  {journal} {\bibinfo
			{journal} {Phys. Rev. A}\ }\textbf {\bibinfo {volume} {93}},\ \bibinfo
		{pages} {033824} (\bibinfo {year} {2016})}\BibitemShut {NoStop}%
	\bibitem [{\citenamefont {Fink}\ \emph {et~al.}(2017)\citenamefont {Fink},
		\citenamefont {Dombi}, \citenamefont {Vukics}, \citenamefont {Wallraff},\
		and\ \citenamefont {Domokos}}]{Fink2017}%
	\BibitemOpen
	\bibfield  {author} {\bibinfo {author} {\bibfnamefont {J.~M.}\ \bibnamefont
			{Fink}}, \bibinfo {author} {\bibfnamefont {A.}~\bibnamefont {Dombi}},
		\bibinfo {author} {\bibfnamefont {A.}~\bibnamefont {Vukics}}, \bibinfo
		{author} {\bibfnamefont {A.}~\bibnamefont {Wallraff}}, \ and\ \bibinfo
		{author} {\bibfnamefont {P.}~\bibnamefont {Domokos}},\ }\href {\doibase
		10.1103/PhysRevX.7.011012} {\bibfield  {journal} {\bibinfo  {journal} {Phys.
				Rev. X}\ }\textbf {\bibinfo {volume} {7}},\ \bibinfo {pages} {011012}
		(\bibinfo {year} {2017})}\BibitemShut {NoStop}%
	\bibitem [{\citenamefont {Fink}\ \emph {et~al.}(2018)\citenamefont {Fink},
		\citenamefont {Schade}, \citenamefont {H{\"o}fling}, \citenamefont
		{Schneider},\ and\ \citenamefont {Imamoglu}}]{Fink2018}%
	\BibitemOpen
	\bibfield  {author} {\bibinfo {author} {\bibfnamefont {T.}~\bibnamefont
			{Fink}}, \bibinfo {author} {\bibfnamefont {A.}~\bibnamefont {Schade}},
		\bibinfo {author} {\bibfnamefont {S.}~\bibnamefont {H{\"o}fling}}, \bibinfo
		{author} {\bibfnamefont {C.}~\bibnamefont {Schneider}}, \ and\ \bibinfo
		{author} {\bibfnamefont {A.}~\bibnamefont {Imamoglu}},\ }\href {\doibase
		10.1038/s41567-017-0020-9} {\bibfield  {journal} {\bibinfo  {journal} {Nature
				Physics}\ }\textbf {\bibinfo {volume} {14}},\ \bibinfo {pages} {365}
		(\bibinfo {year} {2018})}\BibitemShut {NoStop}%
	\bibitem [{\citenamefont {Fitzpatrick}\ \emph {et~al.}(2017)\citenamefont
		{Fitzpatrick}, \citenamefont {Sundaresan}, \citenamefont {Li}, \citenamefont
		{Koch},\ and\ \citenamefont {Houck}}]{Fitzpatrick2017}%
	\BibitemOpen
	\bibfield  {author} {\bibinfo {author} {\bibfnamefont {M.}~\bibnamefont
			{Fitzpatrick}}, \bibinfo {author} {\bibfnamefont {N.~M.}\ \bibnamefont
			{Sundaresan}}, \bibinfo {author} {\bibfnamefont {A.~C.~Y.}\ \bibnamefont
			{Li}}, \bibinfo {author} {\bibfnamefont {J.}~\bibnamefont {Koch}}, \ and\
		\bibinfo {author} {\bibfnamefont {A.~A.}\ \bibnamefont {Houck}},\ }\href
	{\doibase 10.1103/PhysRevX.7.011016} {\bibfield  {journal} {\bibinfo
			{journal} {Phys. Rev. X}\ }\textbf {\bibinfo {volume} {7}},\ \bibinfo {pages}
		{011016} (\bibinfo {year} {2017})}\BibitemShut {NoStop}%
	\bibitem [{\citenamefont {Foss-Feig}\ \emph {et~al.}(2017)\citenamefont
		{Foss-Feig}, \citenamefont {Niroula}, \citenamefont {Young}, \citenamefont
		{Hafezi}, \citenamefont {Gorshkov}, \citenamefont {Wilson},\ and\
		\citenamefont {Maghrebi}}]{Foss-Feig2017}%
	\BibitemOpen
	\bibfield  {author} {\bibinfo {author} {\bibfnamefont {M.}~\bibnamefont
			{Foss-Feig}}, \bibinfo {author} {\bibfnamefont {P.}~\bibnamefont {Niroula}},
		\bibinfo {author} {\bibfnamefont {J.~T.}\ \bibnamefont {Young}}, \bibinfo
		{author} {\bibfnamefont {M.}~\bibnamefont {Hafezi}}, \bibinfo {author}
		{\bibfnamefont {A.~V.}\ \bibnamefont {Gorshkov}}, \bibinfo {author}
		{\bibfnamefont {R.~M.}\ \bibnamefont {Wilson}}, \ and\ \bibinfo {author}
		{\bibfnamefont {M.~F.}\ \bibnamefont {Maghrebi}},\ }\href {\doibase
		10.1103/PhysRevA.95.043826} {\bibfield  {journal} {\bibinfo  {journal} {Phys.
				Rev. A}\ }\textbf {\bibinfo {volume} {95}},\ \bibinfo {pages} {043826}
		(\bibinfo {year} {2017})}\BibitemShut {NoStop}%
	\bibitem [{\citenamefont {Kessler}\ \emph {et~al.}(2012)\citenamefont
		{Kessler}, \citenamefont {Giedke}, \citenamefont {Imamoglu}, \citenamefont
		{Yelin}, \citenamefont {Lukin},\ and\ \citenamefont {Cirac}}]{Kessler2012}%
	\BibitemOpen
	\bibfield  {author} {\bibinfo {author} {\bibfnamefont {E.~M.}\ \bibnamefont
			{Kessler}}, \bibinfo {author} {\bibfnamefont {G.}~\bibnamefont {Giedke}},
		\bibinfo {author} {\bibfnamefont {A.}~\bibnamefont {Imamoglu}}, \bibinfo
		{author} {\bibfnamefont {S.~F.}\ \bibnamefont {Yelin}}, \bibinfo {author}
		{\bibfnamefont {M.~D.}\ \bibnamefont {Lukin}}, \ and\ \bibinfo {author}
		{\bibfnamefont {J.~I.}\ \bibnamefont {Cirac}},\ }\href {\doibase
		10.1103/PhysRevA.86.012116} {\bibfield  {journal} {\bibinfo  {journal} {Phys.
				Rev. A}\ }\textbf {\bibinfo {volume} {86}},\ \bibinfo {pages} {012116}
		(\bibinfo {year} {2012})}\BibitemShut {NoStop}%
	\bibitem [{\citenamefont {Marino}\ and\ \citenamefont
		{Diehl}(2016)}]{Marino2016}%
	\BibitemOpen
	\bibfield  {author} {\bibinfo {author} {\bibfnamefont {J.}~\bibnamefont
			{Marino}}\ and\ \bibinfo {author} {\bibfnamefont {S.}~\bibnamefont {Diehl}},\
	}\href {\doibase 10.1103/PhysRevLett.116.070407} {\bibfield  {journal}
		{\bibinfo  {journal} {Phys. Rev. Lett.}\ }\textbf {\bibinfo {volume} {116}},\
		\bibinfo {pages} {070407} (\bibinfo {year} {2016})}\BibitemShut {NoStop}%
	\bibitem [{\citenamefont {Savona}(2017)}]{Savona2017}%
	\BibitemOpen
	\bibfield  {author} {\bibinfo {author} {\bibfnamefont {V.}~\bibnamefont
			{Savona}},\ }\href {\doibase 10.1103/PhysRevA.96.033826} {\bibfield
		{journal} {\bibinfo  {journal} {Phys. Rev. A}\ }\textbf {\bibinfo {volume}
			{96}},\ \bibinfo {pages} {033826} (\bibinfo {year} {2017})}\BibitemShut
	{NoStop}%
	\bibitem [{\citenamefont {Sieberer}\ \emph {et~al.}(2013)\citenamefont
		{Sieberer}, \citenamefont {Huber}, \citenamefont {Altman},\ and\
		\citenamefont {Diehl}}]{Sieberer2013}%
	\BibitemOpen
	\bibfield  {author} {\bibinfo {author} {\bibfnamefont {L.~M.}\ \bibnamefont
			{Sieberer}}, \bibinfo {author} {\bibfnamefont {S.~D.}\ \bibnamefont {Huber}},
		\bibinfo {author} {\bibfnamefont {E.}~\bibnamefont {Altman}}, \ and\ \bibinfo
		{author} {\bibfnamefont {S.}~\bibnamefont {Diehl}},\ }\href {\doibase
		10.1103/PhysRevLett.110.195301} {\bibfield  {journal} {\bibinfo  {journal}
			{Phys. Rev. Lett.}\ }\textbf {\bibinfo {volume} {110}},\ \bibinfo {pages}
		{195301} (\bibinfo {year} {2013})}\BibitemShut {NoStop}%
	\bibitem [{\citenamefont {Vicentini}\ \emph {et~al.}(2018)\citenamefont
		{Vicentini}, \citenamefont {Minganti}, \citenamefont {Rota}, \citenamefont
		{Orso},\ and\ \citenamefont {Ciuti}}]{Vicentini2018}%
	\BibitemOpen
	\bibfield  {author} {\bibinfo {author} {\bibfnamefont {F.}~\bibnamefont
			{Vicentini}}, \bibinfo {author} {\bibfnamefont {F.}~\bibnamefont {Minganti}},
		\bibinfo {author} {\bibfnamefont {R.}~\bibnamefont {Rota}}, \bibinfo {author}
		{\bibfnamefont {G.}~\bibnamefont {Orso}}, \ and\ \bibinfo {author}
		{\bibfnamefont {C.}~\bibnamefont {Ciuti}},\ }\href {\doibase
		10.1103/PhysRevA.97.013853} {\bibfield  {journal} {\bibinfo  {journal} {Phys.
				Rev. A}\ }\textbf {\bibinfo {volume} {97}},\ \bibinfo {pages} {013853}
		(\bibinfo {year} {2018})}\BibitemShut {NoStop}%
	\bibitem [{\citenamefont {Jin}\ \emph {et~al.}(2016)\citenamefont {Jin},
		\citenamefont {Biella}, \citenamefont {Viyuela}, \citenamefont {Mazza},
		\citenamefont {Keeling}, \citenamefont {Fazio},\ and\ \citenamefont
		{Rossini}}]{Jin2016}%
	\BibitemOpen
	\bibfield  {author} {\bibinfo {author} {\bibfnamefont {J.}~\bibnamefont
			{Jin}}, \bibinfo {author} {\bibfnamefont {A.}~\bibnamefont {Biella}},
		\bibinfo {author} {\bibfnamefont {O.}~\bibnamefont {Viyuela}}, \bibinfo
		{author} {\bibfnamefont {L.}~\bibnamefont {Mazza}}, \bibinfo {author}
		{\bibfnamefont {J.}~\bibnamefont {Keeling}}, \bibinfo {author} {\bibfnamefont
			{R.}~\bibnamefont {Fazio}}, \ and\ \bibinfo {author} {\bibfnamefont
			{D.}~\bibnamefont {Rossini}},\ }\href {\doibase 10.1103/PhysRevX.6.031011}
	{\bibfield  {journal} {\bibinfo  {journal} {Physical Review X}\ }\textbf
		{\bibinfo {volume} {6}},\ \bibinfo {pages} {031011} (\bibinfo {year}
		{2016})}\BibitemShut {NoStop}%
	\bibitem [{\citenamefont {Lee}\ \emph {et~al.}(2013)\citenamefont {Lee},
		\citenamefont {Gopalakrishnan},\ and\ \citenamefont {Lukin}}]{Lee2013}%
	\BibitemOpen
	\bibfield  {author} {\bibinfo {author} {\bibfnamefont {T.~E.}\ \bibnamefont
			{Lee}}, \bibinfo {author} {\bibfnamefont {S.}~\bibnamefont {Gopalakrishnan}},
		\ and\ \bibinfo {author} {\bibfnamefont {M.~D.}\ \bibnamefont {Lukin}},\
	}\href {\doibase 10.1103/PhysRevLett.110.257204} {\bibfield  {journal}
		{\bibinfo  {journal} {Physical Review Letters}\ }\textbf {\bibinfo {volume}
			{110}},\ \bibinfo {pages} {257204} (\bibinfo {year} {2013})}\BibitemShut
	{NoStop}%
	\bibitem [{\citenamefont {Rota}\ \emph {et~al.}(2018)\citenamefont {Rota},
		\citenamefont {Minganti}, \citenamefont {Biella},\ and\ \citenamefont
		{Ciuti}}]{Rota2018}%
	\BibitemOpen
	\bibfield  {author} {\bibinfo {author} {\bibfnamefont {R.}~\bibnamefont
			{Rota}}, \bibinfo {author} {\bibfnamefont {F.}~\bibnamefont {Minganti}},
		\bibinfo {author} {\bibfnamefont {A.}~\bibnamefont {Biella}}, \ and\ \bibinfo
		{author} {\bibfnamefont {C.}~\bibnamefont {Ciuti}},\ }\href {\doibase
		10.1088/1367-2630/aab703} {\bibfield  {journal} {\bibinfo  {journal} {New
				Journal of Physics}\ }\textbf {\bibinfo {volume} {20}},\ \bibinfo {pages}
		{045003} (\bibinfo {year} {2018})}\BibitemShut {NoStop}%
	\bibitem [{\citenamefont {Rota}\ \emph {et~al.}(2017)\citenamefont {Rota},
		\citenamefont {Storme}, \citenamefont {Bartolo}, \citenamefont {Fazio},\ and\
		\citenamefont {Ciuti}}]{Rota2017}%
	\BibitemOpen
	\bibfield  {author} {\bibinfo {author} {\bibfnamefont {R.}~\bibnamefont
			{Rota}}, \bibinfo {author} {\bibfnamefont {F.}~\bibnamefont {Storme}},
		\bibinfo {author} {\bibfnamefont {N.}~\bibnamefont {Bartolo}}, \bibinfo
		{author} {\bibfnamefont {R.}~\bibnamefont {Fazio}}, \ and\ \bibinfo {author}
		{\bibfnamefont {C.}~\bibnamefont {Ciuti}},\ }\href {\doibase
		10.1103/PhysRevB.95.134431} {\bibfield  {journal} {\bibinfo  {journal}
			{Physical Review B}\ }\textbf {\bibinfo {volume} {95}},\ \bibinfo {pages}
		{134431} (\bibinfo {year} {2017})}\BibitemShut {NoStop}%
	\bibitem [{\citenamefont {Casteels}\ \emph {et~al.}(2018)\citenamefont
		{Casteels}, \citenamefont {Wilson},\ and\ \citenamefont
		{Wouters}}]{Casteels2018}%
	\BibitemOpen
	\bibfield  {author} {\bibinfo {author} {\bibfnamefont {W.}~\bibnamefont
			{Casteels}}, \bibinfo {author} {\bibfnamefont {R.~M.}\ \bibnamefont
			{Wilson}}, \ and\ \bibinfo {author} {\bibfnamefont {M.}~\bibnamefont
			{Wouters}},\ }\href {\doibase 10.1103/PhysRevA.97.062107} {\bibfield
		{journal} {\bibinfo  {journal} {Physical Review A}\ }\textbf {\bibinfo
			{volume} {97}},\ \bibinfo {pages} {062107} (\bibinfo {year}
		{2018})}\BibitemShut {NoStop}%
	\bibitem [{\citenamefont {Bremner}\ \emph {et~al.}(2016)\citenamefont
		{Bremner}, \citenamefont {Montanaro},\ and\ \citenamefont
		{Shepherd}}]{Bremner2016}%
	\BibitemOpen
	\bibfield  {author} {\bibinfo {author} {\bibfnamefont {M.~J.}\ \bibnamefont
			{Bremner}}, \bibinfo {author} {\bibfnamefont {A.}~\bibnamefont {Montanaro}},
		\ and\ \bibinfo {author} {\bibfnamefont {D.~J.}\ \bibnamefont {Shepherd}},\
	}\href {\doibase 10.22331/q-2017-04-25-8} {\bibfield  {journal} {\bibinfo
			{journal} {Quantum}\ }\textbf {\bibinfo {volume} {1}},\ \bibinfo {pages} {8}
		(\bibinfo {year} {2016})},\ \Eprint {http://arxiv.org/abs/1610.01808}
	{arXiv:1610.01808} \BibitemShut {NoStop}%
	\bibitem [{\citenamefont {Gao}\ and\ \citenamefont {Duan}(2018)}]{Gao2018}%
	\BibitemOpen
	\bibfield  {author} {\bibinfo {author} {\bibfnamefont {X.}~\bibnamefont
			{Gao}}\ and\ \bibinfo {author} {\bibfnamefont {L.}~\bibnamefont {Duan}},\
	}\href {http://arxiv.org/abs/1810.03176} {\  (\bibinfo {year} {2018})},\
	\Eprint {http://arxiv.org/abs/1810.03176} {arXiv:1810.03176} \BibitemShut
	{NoStop}%
	\bibitem [{\citenamefont {Harrow}\ and\ \citenamefont
		{Montanaro}(2017)}]{Harrow2017}%
	\BibitemOpen
	\bibfield  {author} {\bibinfo {author} {\bibfnamefont {A.~W.}\ \bibnamefont
			{Harrow}}\ and\ \bibinfo {author} {\bibfnamefont {A.}~\bibnamefont
			{Montanaro}},\ }\href {\doibase 10.1038/nature23458} {\bibfield  {journal}
		{\bibinfo  {journal} {Nature}\ }\textbf {\bibinfo {volume} {549}},\ \bibinfo
		{pages} {203} (\bibinfo {year} {2017})}\BibitemShut {NoStop}%
	\bibitem [{\citenamefont {Preskill}(2018)}]{Preskill2018}%
	\BibitemOpen
	\bibfield  {author} {\bibinfo {author} {\bibfnamefont {J.}~\bibnamefont
			{Preskill}},\ }\href {\doibase 10.22331/q-2018-08-06-79} {\bibfield
		{journal} {\bibinfo  {journal} {Quantum}\ }\textbf {\bibinfo {volume} {2}},\
		\bibinfo {pages} {79} (\bibinfo {year} {2018})},\ \Eprint
	{http://arxiv.org/abs/1801.00862} {arXiv:1801.00862} \BibitemShut {NoStop}%
	\bibitem [{\citenamefont {Breuer}\ and\ \citenamefont
		{Petruccione}(2002)}]{Breuer2002}%
	\BibitemOpen
	\bibfield  {author} {\bibinfo {author} {\bibfnamefont {H.-P.}\ \bibnamefont
			{Breuer}}\ and\ \bibinfo {author} {\bibfnamefont {F.~F.}\ \bibnamefont
			{Petruccione}},\ }\href
	{https://books.google.ch/books/about/The{\_}Theory{\_}of{\_}Open{\_}Quantum{\_}Systems.html?id=0Yx5VzaMYm8C{\&}redir{\_}esc=y}
	{\emph {\bibinfo {title} {{The theory of open quantum systems}}}}\ (\bibinfo
	{publisher} {Oxford University Press},\ \bibinfo {year} {2002})\ p.\ \bibinfo
	{pages} {625}\BibitemShut {NoStop}%
	\bibitem [{\citenamefont {Prosen}(2011)}]{Prosen2011}%
	\BibitemOpen
	\bibfield  {author} {\bibinfo {author} {\bibfnamefont {T.~c.~v.}\
			\bibnamefont {Prosen}},\ }\href {\doibase 10.1103/PhysRevLett.107.137201}
	{\bibfield  {journal} {\bibinfo  {journal} {Phys. Rev. Lett.}\ }\textbf
		{\bibinfo {volume} {107}},\ \bibinfo {pages} {137201} (\bibinfo {year}
		{2011})}\BibitemShut {NoStop}%
	\bibitem [{\citenamefont {Prosen}(2014)}]{Prosen2014}%
	\BibitemOpen
	\bibfield  {author} {\bibinfo {author} {\bibfnamefont {T.~c.~v.}\
			\bibnamefont {Prosen}},\ }\href {\doibase 10.1103/PhysRevLett.112.030603}
	{\bibfield  {journal} {\bibinfo  {journal} {Phys. Rev. Lett.}\ }\textbf
		{\bibinfo {volume} {112}},\ \bibinfo {pages} {030603} (\bibinfo {year}
		{2014})}\BibitemShut {NoStop}%
	\bibitem [{\citenamefont {Cui}\ \emph {et~al.}(2015)\citenamefont {Cui},
		\citenamefont {Cirac},\ and\ \citenamefont {Ba{\~{n}}uls}}]{Cui2015}%
	\BibitemOpen
	\bibfield  {author} {\bibinfo {author} {\bibfnamefont {J.}~\bibnamefont
			{Cui}}, \bibinfo {author} {\bibfnamefont {J.~I.}\ \bibnamefont {Cirac}}, \
		and\ \bibinfo {author} {\bibfnamefont {M.~C.}\ \bibnamefont {Ba{\~{n}}uls}},\
	}\href {\doibase 10.1103/PhysRevLett.114.220601} {\bibfield  {journal}
		{\bibinfo  {journal} {Physical Review Letters}\ }\textbf {\bibinfo {volume}
			{114}},\ \bibinfo {pages} {220601} (\bibinfo {year} {2015})}\BibitemShut
	{NoStop}%
	\bibitem [{\citenamefont {Kshetrimayum}\ \emph {et~al.}(2017)\citenamefont
		{Kshetrimayum}, \citenamefont {Weimer},\ and\ \citenamefont
		{Or{\'{u}}s}}]{Kshetrimayum2017}%
	\BibitemOpen
	\bibfield  {author} {\bibinfo {author} {\bibfnamefont {A.}~\bibnamefont
			{Kshetrimayum}}, \bibinfo {author} {\bibfnamefont {H.}~\bibnamefont
			{Weimer}}, \ and\ \bibinfo {author} {\bibfnamefont {R.}~\bibnamefont
			{Or{\'{u}}s}},\ }\href {\doibase 10.1038/s41467-017-01511-6} {\bibfield
		{journal} {\bibinfo  {journal} {Nature Communications}\ }\textbf {\bibinfo
			{volume} {8}},\ \bibinfo {pages} {1291} (\bibinfo {year} {2017})}\BibitemShut
	{NoStop}%
	\bibitem [{\citenamefont {Mascarenhas}\ \emph {et~al.}(2015)\citenamefont
		{Mascarenhas}, \citenamefont {Flayac},\ and\ \citenamefont
		{Savona}}]{Mascarenhas2015}%
	\BibitemOpen
	\bibfield  {author} {\bibinfo {author} {\bibfnamefont {E.}~\bibnamefont
			{Mascarenhas}}, \bibinfo {author} {\bibfnamefont {H.}~\bibnamefont {Flayac}},
		\ and\ \bibinfo {author} {\bibfnamefont {V.}~\bibnamefont {Savona}},\ }\href
	{\doibase 10.1103/PhysRevA.92.022116} {\bibfield  {journal} {\bibinfo
			{journal} {Physical Review A - Atomic, Molecular, and Optical Physics}\
		}\textbf {\bibinfo {volume} {92}},\ \bibinfo {pages} {56} (\bibinfo {year}
		{2015})},\ \Eprint {http://arxiv.org/abs/1504.06127} {arXiv:1504.06127}
	\BibitemShut {NoStop}%
	\bibitem [{\citenamefont {Werner}\ \emph {et~al.}(2016)\citenamefont {Werner},
		\citenamefont {Jaschke}, \citenamefont {Silvi}, \citenamefont {Kliesch},
		\citenamefont {Calarco}, \citenamefont {Eisert},\ and\ \citenamefont
		{Montangero}}]{Werner2016}%
	\BibitemOpen
	\bibfield  {author} {\bibinfo {author} {\bibfnamefont {A.}~\bibnamefont
			{Werner}}, \bibinfo {author} {\bibfnamefont {D.}~\bibnamefont {Jaschke}},
		\bibinfo {author} {\bibfnamefont {P.}~\bibnamefont {Silvi}}, \bibinfo
		{author} {\bibfnamefont {M.}~\bibnamefont {Kliesch}}, \bibinfo {author}
		{\bibfnamefont {T.}~\bibnamefont {Calarco}}, \bibinfo {author} {\bibfnamefont
			{J.}~\bibnamefont {Eisert}}, \ and\ \bibinfo {author} {\bibfnamefont
			{S.}~\bibnamefont {Montangero}},\ }\href {\doibase
		10.1103/PhysRevLett.116.237201} {\bibfield  {journal} {\bibinfo  {journal}
			{Physical Review Letters}\ }\textbf {\bibinfo {volume} {116}},\ \bibinfo
		{pages} {237201} (\bibinfo {year} {2016})}\BibitemShut {NoStop}%
	\bibitem [{\citenamefont {Finazzi}\ \emph {et~al.}(2015)\citenamefont
		{Finazzi}, \citenamefont {{Le Boit{\'{e}}}}, \citenamefont {Storme},
		\citenamefont {Baksic},\ and\ \citenamefont {Ciuti}}]{Finazzi2015}%
	\BibitemOpen
	\bibfield  {author} {\bibinfo {author} {\bibfnamefont {S.}~\bibnamefont
			{Finazzi}}, \bibinfo {author} {\bibfnamefont {A.}~\bibnamefont {{Le
					Boit{\'{e}}}}}, \bibinfo {author} {\bibfnamefont {F.}~\bibnamefont {Storme}},
		\bibinfo {author} {\bibfnamefont {A.}~\bibnamefont {Baksic}}, \ and\ \bibinfo
		{author} {\bibfnamefont {C.}~\bibnamefont {Ciuti}},\ }\href {\doibase
		10.1103/PhysRevLett.115.080604} {\bibfield  {journal} {\bibinfo  {journal}
			{Physical Review Letters}\ }\textbf {\bibinfo {volume} {115}},\ \bibinfo
		{pages} {080604} (\bibinfo {year} {2015})}\BibitemShut {NoStop}%
	\bibitem [{\citenamefont {Nagy}\ and\ \citenamefont
		{Savona}(2018)}]{nagy_driven-dissipative_2018}%
	\BibitemOpen
	\bibfield  {author} {\bibinfo {author} {\bibfnamefont {A.}~\bibnamefont
			{Nagy}}\ and\ \bibinfo {author} {\bibfnamefont {V.}~\bibnamefont {Savona}},\
	}\href {\doibase 10.1103/PhysRevA.97.052129} {\bibfield  {journal} {\bibinfo
			{journal} {Phys. Rev. A}\ }\textbf {\bibinfo {volume} {97}},\ \bibinfo
		{pages} {052129} (\bibinfo {year} {2018})}\BibitemShut {NoStop}%
	\bibitem [{\citenamefont {Cai}\ and\ \citenamefont {Liu}(2018)}]{Cai2018}%
	\BibitemOpen
	\bibfield  {author} {\bibinfo {author} {\bibfnamefont {Z.}~\bibnamefont
			{Cai}}\ and\ \bibinfo {author} {\bibfnamefont {J.}~\bibnamefont {Liu}},\
	}\href {\doibase 10.1103/PhysRevB.97.035116} {\bibfield  {journal} {\bibinfo
			{journal} {Physical Review B}\ }\textbf {\bibinfo {volume} {97}},\ \bibinfo
		{pages} {035116} (\bibinfo {year} {2018})}\BibitemShut {NoStop}%
	\bibitem [{\citenamefont {Carleo}\ \emph {et~al.}(2018)\citenamefont {Carleo},
		\citenamefont {Nomura},\ and\ \citenamefont {Imada}}]{Carleo2018}%
	\BibitemOpen
	\bibfield  {author} {\bibinfo {author} {\bibfnamefont {G.}~\bibnamefont
			{Carleo}}, \bibinfo {author} {\bibfnamefont {Y.}~\bibnamefont {Nomura}}, \
		and\ \bibinfo {author} {\bibfnamefont {M.}~\bibnamefont {Imada}},\ }\href
	{\doibase 10.1038/s41467-018-07520-3} {\bibfield  {journal} {\bibinfo
			{journal} {Nature Communications}\ }\textbf {\bibinfo {volume} {9}},\
		\bibinfo {pages} {5322} (\bibinfo {year} {2018})}\BibitemShut {NoStop}%
	\bibitem [{\citenamefont {Carleo}\ and\ \citenamefont
		{Troyer}(2017)}]{Carleo2017}%
	\BibitemOpen
	\bibfield  {author} {\bibinfo {author} {\bibfnamefont {G.}~\bibnamefont
			{Carleo}}\ and\ \bibinfo {author} {\bibfnamefont {M.}~\bibnamefont
			{Troyer}},\ }\href {\doibase 10.1126/science.aag2302} {\bibfield  {journal}
		{\bibinfo  {journal} {Science}\ }\textbf {\bibinfo {volume} {355}},\ \bibinfo
		{pages} {602} (\bibinfo {year} {2017})}\BibitemShut {NoStop}%
	\bibitem [{\citenamefont {Freitas}\ \emph {et~al.}(2018)\citenamefont
		{Freitas}, \citenamefont {Morigi},\ and\ \citenamefont
		{Dunjko}}]{Freitas2018}%
	\BibitemOpen
	\bibfield  {author} {\bibinfo {author} {\bibfnamefont {N.}~\bibnamefont
			{Freitas}}, \bibinfo {author} {\bibfnamefont {G.}~\bibnamefont {Morigi}}, \
		and\ \bibinfo {author} {\bibfnamefont {V.}~\bibnamefont {Dunjko}},\ }\href
	{\doibase 10.1142/S0219749918400087} {\bibfield  {journal} {\bibinfo
			{journal} {International Journal of Quantum Information}\ }\textbf {\bibinfo
			{volume} {16}},\ \bibinfo {pages} {1840008} (\bibinfo {year}
		{2018})}\BibitemShut {NoStop}%
	\bibitem [{\citenamefont {Glasser}\ \emph {et~al.}(2018)\citenamefont
		{Glasser}, \citenamefont {Pancotti}, \citenamefont {August}, \citenamefont
		{Rodriguez},\ and\ \citenamefont {Cirac}}]{Glasser2018}%
	\BibitemOpen
	\bibfield  {author} {\bibinfo {author} {\bibfnamefont {I.}~\bibnamefont
			{Glasser}}, \bibinfo {author} {\bibfnamefont {N.}~\bibnamefont {Pancotti}},
		\bibinfo {author} {\bibfnamefont {M.}~\bibnamefont {August}}, \bibinfo
		{author} {\bibfnamefont {I.~D.}\ \bibnamefont {Rodriguez}}, \ and\ \bibinfo
		{author} {\bibfnamefont {J.~I.}\ \bibnamefont {Cirac}},\ }\href {\doibase
		10.1103/PhysRevX.8.011006} {\bibfield  {journal} {\bibinfo  {journal}
			{Physical Review X}\ }\textbf {\bibinfo {volume} {8}},\ \bibinfo {pages}
		{011006} (\bibinfo {year} {2018})}\BibitemShut {NoStop}%
	\bibitem [{\citenamefont {Nomura}\ \emph {et~al.}(2017)\citenamefont {Nomura},
		\citenamefont {Darmawan}, \citenamefont {Yamaji},\ and\ \citenamefont
		{Imada}}]{Nomura2017}%
	\BibitemOpen
	\bibfield  {author} {\bibinfo {author} {\bibfnamefont {Y.}~\bibnamefont
			{Nomura}}, \bibinfo {author} {\bibfnamefont {A.~S.}\ \bibnamefont
			{Darmawan}}, \bibinfo {author} {\bibfnamefont {Y.}~\bibnamefont {Yamaji}}, \
		and\ \bibinfo {author} {\bibfnamefont {M.}~\bibnamefont {Imada}},\ }\href
	{\doibase 10.1103/PhysRevB.96.205152} {\bibfield  {journal} {\bibinfo
			{journal} {Phys. Rev. B}\ }\textbf {\bibinfo {volume} {96}},\ \bibinfo
		{pages} {205152} (\bibinfo {year} {2017})}\BibitemShut {NoStop}%
	\bibitem [{\citenamefont {Gao}\ and\ \citenamefont {Duan}(2017)}]{Gao2017}%
	\BibitemOpen
	\bibfield  {author} {\bibinfo {author} {\bibfnamefont {X.}~\bibnamefont
			{Gao}}\ and\ \bibinfo {author} {\bibfnamefont {L.-M.}\ \bibnamefont {Duan}},\
	}\href {https://doi.org/10.1038/s41467-017-00705-2} {\bibfield  {journal}
		{\bibinfo  {journal} {Nature Communications}\ }\textbf {\bibinfo {volume}
			{8}},\ \bibinfo {pages} {662} (\bibinfo {year} {2017})}\BibitemShut {NoStop}%
	\bibitem [{\citenamefont {Torlai}\ and\ \citenamefont
		{Melko}(2018)}]{Torlai2018}%
	\BibitemOpen
	\bibfield  {author} {\bibinfo {author} {\bibfnamefont {G.}~\bibnamefont
			{Torlai}}\ and\ \bibinfo {author} {\bibfnamefont {R.~G.}\ \bibnamefont
			{Melko}},\ }\href {\doibase 10.1103/PhysRevLett.120.240503} {\bibfield
		{journal} {\bibinfo  {journal} {Phys. Rev. Lett.}\ }\textbf {\bibinfo
			{volume} {120}},\ \bibinfo {pages} {240503} (\bibinfo {year}
		{2018})}\BibitemShut {NoStop}%
	\bibitem [{\citenamefont {Jakob}\ and\ \citenamefont
		{Stenholm}(2003)}]{Jakob2003}%
	\BibitemOpen
	\bibfield  {author} {\bibinfo {author} {\bibfnamefont {M.}~\bibnamefont
			{Jakob}}\ and\ \bibinfo {author} {\bibfnamefont {S.}~\bibnamefont
			{Stenholm}},\ }\href {\doibase 10.1103/PhysRevA.67.032111} {\bibfield
		{journal} {\bibinfo  {journal} {Phys. Rev. A}\ }\textbf {\bibinfo {volume}
			{67}},\ \bibinfo {pages} {032111} (\bibinfo {year} {2003})}\BibitemShut
	{NoStop}%
	\bibitem [{\citenamefont {Weimer}(2015)}]{Weimer2015}%
	\BibitemOpen
	\bibfield  {author} {\bibinfo {author} {\bibfnamefont {H.}~\bibnamefont
			{Weimer}},\ }\href {\doibase 10.1103/PhysRevLett.114.040402} {\bibfield
		{journal} {\bibinfo  {journal} {Physical Review Letters}\ }\textbf {\bibinfo
			{volume} {114}},\ \bibinfo {pages} {040402} (\bibinfo {year}
		{2015})}\BibitemShut {NoStop}%
	\bibitem [{\citenamefont {Sorella}\ \emph {et~al.}(2007)\citenamefont
		{Sorella}, \citenamefont {Casula},\ and\ \citenamefont
		{Rocca}}]{Sorella2007}%
	\BibitemOpen
	\bibfield  {author} {\bibinfo {author} {\bibfnamefont {S.}~\bibnamefont
			{Sorella}}, \bibinfo {author} {\bibfnamefont {M.}~\bibnamefont {Casula}}, \
		and\ \bibinfo {author} {\bibfnamefont {D.}~\bibnamefont {Rocca}},\ }\href
	{\doibase 10.1063/1.2746035} {\bibfield  {journal} {\bibinfo  {journal} {J.
				Chem. Phys}\ }\textbf {\bibinfo {volume} {127}},\ \bibinfo {pages} {014105}
		(\bibinfo {year} {2007})}\BibitemShut {NoStop}%
	\bibitem [{\citenamefont {Nigro}(2018)}]{Nigro2018}%
	\BibitemOpen
	\bibfield  {author} {\bibinfo {author} {\bibfnamefont {D.}~\bibnamefont
			{Nigro}},\ }\href {http://arxiv.org/abs/1803.06279} {\  (\bibinfo {year}
		{2018})},\ \Eprint {http://arxiv.org/abs/1803.06279} {arXiv:1803.06279}
	\BibitemShut {NoStop}%
	\bibitem [{\citenamefont {Minganti}\ \emph {et~al.}(2018)\citenamefont
		{Minganti}, \citenamefont {Biella}, \citenamefont {Bartolo},\ and\
		\citenamefont {Ciuti}}]{Minganti2018}%
	\BibitemOpen
	\bibfield  {author} {\bibinfo {author} {\bibfnamefont {F.}~\bibnamefont
			{Minganti}}, \bibinfo {author} {\bibfnamefont {A.}~\bibnamefont {Biella}},
		\bibinfo {author} {\bibfnamefont {N.}~\bibnamefont {Bartolo}}, \ and\
		\bibinfo {author} {\bibfnamefont {C.}~\bibnamefont {Ciuti}},\ }\href
	{\doibase 10.1103/PhysRevA.98.042118} {\bibfield  {journal} {\bibinfo
			{journal} {Physical Review A}\ }\textbf {\bibinfo {volume} {98}},\ \bibinfo
		{pages} {042118} (\bibinfo {year} {2018})}\BibitemShut {NoStop}%
	\bibitem [{sup()}]{supplemental}%
	\BibitemOpen
	\href@noop {} {\ }\bibinfo {note} {See Supplemental Material for more information on the neural network ansatz, the
		stochastic reconfiguration method, the stochastic sampling procedure and the
		details of the numerical approach, which includes Refs. \cite{Carleo2017,Sorella2007,metropolis_equation_1953,choi_minres-qlp:_2011,liu_contribute_2019}.}\BibitemShut {Stop}%
	
	\bibitem [{\citenamefont {Choi}\ \emph {et~al.}(2011)\citenamefont {Choi},
		\citenamefont {Paige},\ and\ \citenamefont
		{Saunders}}]{choi_minres-qlp:_2011}%
	\BibitemOpen
	\bibfield  {author} {\bibinfo {author} {\bibfnamefont {S.}~\bibnamefont
			{Choi}}, \bibinfo {author} {\bibfnamefont {C.}~\bibnamefont {Paige}}, \ and\
		\bibinfo {author} {\bibfnamefont {M.}~\bibnamefont {Saunders}},\ }\href
	{\doibase 10.1137/100787921} {\bibfield  {journal} {\bibinfo  {journal} {SIAM
				J. Sci. Comput.}\ }\textbf {\bibinfo {volume} {33}},\ \bibinfo {pages} {1810}
		(\bibinfo {year} {2011})}\BibitemShut {NoStop}%
	\bibitem [{\citenamefont {Liu}(2019)}]{liu_contribute_2019}%
	\BibitemOpen
	\bibfield  {author} {\bibinfo {author} {\bibfnamefont {Y.}~\bibnamefont
			{Liu}},\ }\href {https://github.com/syangliu/MINRES-QLP} {\enquote {\bibinfo
			{title} {Contribute to syangliu/{MINRES}-{QLP} development by creating an
				account on {GitHub}},}\ } (\bibinfo {year} {2019}),\ \bibinfo {note}
	{original-date: 2018-06-08T04:18:21Z}\BibitemShut {NoStop}%
	
	\bibitem [{\citenamefont {Le~Roux}\ and\ \citenamefont
		{Bengio}(2008)}]{Leroux2008}%
	\BibitemOpen
	\bibfield  {author} {\bibinfo {author} {\bibfnamefont {N.}~\bibnamefont
			{Le~Roux}}\ and\ \bibinfo {author} {\bibfnamefont {Y.}~\bibnamefont
			{Bengio}},\ }\href {\doibase 10.1162/neco.2008.04-07-510} {\bibfield
		{journal} {\bibinfo  {journal} {Neural Computation}\ }\textbf {\bibinfo
			{volume} {20}},\ \bibinfo {pages} {1631} (\bibinfo {year} {2008})},\ \Eprint
	{http://arxiv.org/abs/https://doi.org/10.1162/neco.2008.04-07-510}
	{https://doi.org/10.1162/neco.2008.04-07-510} \BibitemShut {NoStop}%
	\bibitem [{\citenamefont {Torlai}\ \emph {et~al.}(2018)\citenamefont {Torlai},
		\citenamefont {Mazzola}, \citenamefont {Carrasquilla}, \citenamefont
		{Troyer}, \citenamefont {Melko},\ and\ \citenamefont {Carleo}}]{Torlai2018a}%
	\BibitemOpen
	\bibfield  {author} {\bibinfo {author} {\bibfnamefont {G.}~\bibnamefont
			{Torlai}}, \bibinfo {author} {\bibfnamefont {G.}~\bibnamefont {Mazzola}},
		\bibinfo {author} {\bibfnamefont {J.}~\bibnamefont {Carrasquilla}}, \bibinfo
		{author} {\bibfnamefont {M.}~\bibnamefont {Troyer}}, \bibinfo {author}
		{\bibfnamefont {R.}~\bibnamefont {Melko}}, \ and\ \bibinfo {author}
		{\bibfnamefont {G.}~\bibnamefont {Carleo}},\ }\href {\doibase
		10.1038/s41567-018-0048-5} {\bibfield  {journal} {\bibinfo  {journal} {Nature
				Physics}\ }\textbf {\bibinfo {volume} {14}},\ \bibinfo {pages} {447}
		(\bibinfo {year} {2018})}\BibitemShut {NoStop}%
	\bibitem [{\citenamefont {Metropolis}\ \emph {et~al.}(1953)\citenamefont
		{Metropolis}, \citenamefont {Rosenbluth}, \citenamefont {Rosenbluth},
		\citenamefont {Teller},\ and\ \citenamefont
		{Teller}}]{metropolis_equation_1953}%
	\BibitemOpen
	\bibfield  {author} {\bibinfo {author} {\bibfnamefont {N.}~\bibnamefont
			{Metropolis}}, \bibinfo {author} {\bibfnamefont {A.~W.}\ \bibnamefont
			{Rosenbluth}}, \bibinfo {author} {\bibfnamefont {M.~N.}\ \bibnamefont
			{Rosenbluth}}, \bibinfo {author} {\bibfnamefont {A.~H.}\ \bibnamefont
			{Teller}}, \ and\ \bibinfo {author} {\bibfnamefont {E.}~\bibnamefont
			{Teller}},\ }\href {\doibase 10.1063/1.1699114} {\bibfield  {journal}
		{\bibinfo  {journal} {J. Chem. Phys}\ }\textbf {\bibinfo {volume} {21}},\
		\bibinfo {pages} {1087} (\bibinfo {year} {1953})}\BibitemShut {NoStop}%
	\bibitem [{\citenamefont {Choo}\ \emph {et~al.}(2018)\citenamefont {Choo},
		\citenamefont {Carleo}, \citenamefont {Regnault},\ and\ \citenamefont
		{Neupert}}]{Choo2018}%
	\BibitemOpen
	\bibfield  {author} {\bibinfo {author} {\bibfnamefont {K.}~\bibnamefont
			{Choo}}, \bibinfo {author} {\bibfnamefont {G.}~\bibnamefont {Carleo}},
		\bibinfo {author} {\bibfnamefont {N.}~\bibnamefont {Regnault}}, \ and\
		\bibinfo {author} {\bibfnamefont {T.}~\bibnamefont {Neupert}},\ }\href
	{\doibase 10.1103/PhysRevLett.121.167204} {\bibfield  {journal} {\bibinfo
			{journal} {Phys. Rev. Lett.}\ }\textbf {\bibinfo {volume} {121}},\ \bibinfo
		{pages} {167204} (\bibinfo {year} {2018})}\BibitemShut {NoStop}%
	\bibitem [{\citenamefont {Saito}(2017)}]{Saito2017}%
	\BibitemOpen
	\bibfield  {author} {\bibinfo {author} {\bibfnamefont {H.}~\bibnamefont
			{Saito}},\ }\href {\doibase 10.7566/JPSJ.86.093001} {\bibfield  {journal}
		{\bibinfo  {journal} {Journal of the Physical Society of Japan}\ }\textbf
		{\bibinfo {volume} {86}},\ \bibinfo {pages} {093001} (\bibinfo {year}
		{2017})}\BibitemShut {NoStop}%
	\bibitem [{\citenamefont {Hartmann}\ and\ \citenamefont
		{Carleo}(2019)}]{hartmann_neural-network_2019}%
	\BibitemOpen
	\bibfield  {author} {\bibinfo {author} {\bibfnamefont {M.~J.}\ \bibnamefont
			{Hartmann}}\ and\ \bibinfo {author} {\bibfnamefont {G.}~\bibnamefont
			{Carleo}},\ }\href {http://arxiv.org/abs/1902.05131} {\bibfield  {journal}
		{\bibinfo  {journal} {arXiv:1902.05131 [cond-mat, physics:quant-ph]}\ }
		(\bibinfo {year} {2019})},\ \bibinfo {note} {arXiv: 1902.05131}\BibitemShut
	{NoStop}%
	\bibitem [{\citenamefont {Vicentini}\ \emph {et~al.}(2019)\citenamefont
		{Vicentini}, \citenamefont {Biella}, \citenamefont {Regnault},\ and\
		\citenamefont {Ciuti}}]{vicentini_variational_2019}%
	\BibitemOpen
	\bibfield  {author} {\bibinfo {author} {\bibfnamefont {F.}~\bibnamefont
			{Vicentini}}, \bibinfo {author} {\bibfnamefont {A.}~\bibnamefont {Biella}},
		\bibinfo {author} {\bibfnamefont {N.}~\bibnamefont {Regnault}}, \ and\
		\bibinfo {author} {\bibfnamefont {C.}~\bibnamefont {Ciuti}},\ }\href
	{http://arxiv.org/abs/1902.10104} {\bibfield  {journal} {\bibinfo  {journal}
			{arXiv:1902.10104 [cond-mat, physics:quant-ph]}\ } (\bibinfo {year}
		{2019})},\ \bibinfo {note} {arXiv: 1902.10104}\BibitemShut {NoStop}%
	\bibitem [{\citenamefont {Yoshioka}\ and\ \citenamefont
		{Hamazaki}(2019)}]{yoshioka_constructing_2019}%
	\BibitemOpen
	\bibfield  {author} {\bibinfo {author} {\bibfnamefont {N.}~\bibnamefont
			{Yoshioka}}\ and\ \bibinfo {author} {\bibfnamefont {R.}~\bibnamefont
			{Hamazaki}},\ }\href {http://arxiv.org/abs/1902.07006} {\bibfield  {journal}
		{\bibinfo  {journal} {arXiv:1902.07006 [cond-mat, physics:quant-ph]}\ }
		(\bibinfo {year} {2019})},\ \bibinfo {note} {arXiv: 1902.07006}\BibitemShut
	{NoStop}%
\end{thebibliography}
\end{document}